\pdfoutput=1
\documentclass[fleqn,usenatbib]{mnras}

\usepackage[T1]{fontenc}
\usepackage{ae,aecompl}

\usepackage{graphicx}	
\usepackage{amsmath}	
\usepackage{amssymb}	
\usepackage{multirow}

\newcommand{\Eq}[1]{Eq.~\ref{#1}}
\newcommand{\Fig}[1]{Fig.~\ref{#1}}
\newcommand{\Sec}[1]{Section~\ref{#1}}
\newcommand{\App}[1]{Appendix~\ref{#1}}
\newcommand{\Tab}[1]{Table~\ref{#1}}
\newcommand{\threehun}{{\sc TheThreeHundred}}

\title[Contrasting simulations of galaxy groups]{The Three Hundred Project: Substructure in hydrodynamical and dark matter simulations of galaxy groups around clusters}
\author[R. Haggar et al.]{\parbox{\textwidth}{
Roan Haggar,$^{1}$\thanks{E-mail: roan.haggar@nottingham.ac.uk}
Frazer R. Pearce,$^{1}$
Meghan E. Gray,$^{1}$
Alexander Knebe,$^{2,3,4}$}
\newauthor{Gustavo Yepes,$^{2,3}$
}
\\\\
\parbox{\textwidth}{
$^{1}$School and Physics \& Astronomy, University of Nottingham, Nottingham NG7 2RD, UK\\
$^{2}$Departamento de F\'isica Te\'{o}rica, M\'{o}dulo 15 Universidad Aut\'{o}noma de Madrid, 28049 Madrid, Spain\\
$^{3}$Centro de Investigaci\'{o}n Avanzada en F\'{\i}sica Fundamental (CIAFF), Universidad Aut\'{o}noma de Madrid, 28049 Madrid, Spain\\
$^{4}$International Centre for Radio Astronomy Research, The University of Western Australia, 35 Stirling Highway, Crawley, Western Australia 6009, Australia
}}

\date{Accepted 2020 December 22. Received 2020 December 15; in original form 2020 September 18}

\pubyear{2021}

\begin{document}
\label{firstpage}
\pagerange{\pageref{firstpage}--\pageref{lastpage}}
\maketitle

\begin{abstract}
Dark matter-only simulations are able to produce the cosmic structure of a $\Lambda$CDM universe, at a much lower computational cost than more physically motivated hydrodynamical simulations. However, it is not clear how well smaller substructure is reproduced by dark matter-only simulations. To investigate this, we directly compare the substructure of galaxy clusters and of surrounding galaxy groups in hydrodynamical and dark matter-only simulations. We utilise \threehun\ project, a suite of 324 simulations of galaxy clusters that have been simulated with hydrodynamics, and in dark matter-only. We find that dark matter-only simulations underestimate the number density of galaxies in the centres of groups and clusters relative to hydrodynamical simulations, and that this effect is stronger in denser regions. We also look at the phase space of infalling galaxy groups, to show that dark matter-only simulations underpredict the number density of galaxies in the centres of these groups by about a factor of four. This implies that the structure and evolution of infalling groups may be different to that predicted by dark matter-only simulations. Finally, we discuss potential causes for this underestimation, considering both physical effects, and numerical differences in the analysis. 
\end{abstract}




\section{Introduction}
\label{sec:intro}

Cosmological simulations are a valuable tool in testing predictions of the $\Lambda$ cold dark matter ($\Lambda$CDM) model of the Universe, widely accepted as the `standard model' of cosmology \citep{diemand2011, frenk2012}. In this paradigm, small haloes form via the gravitational collapse of cold dark matter. These then grow hierarchically, merging with other haloes to form increasingly massive bound structures, the largest of which are galaxy clusters \citep{white1978}.

The mass composition of galaxy clusters is dominated by dark matter, which makes up over $80\%$ of the mass of a typical cluster \citep{allen2011}, and so the gravitational collapse of these structures is dominated by the effects of dark matter \citep{jenkins1998, springel2005b, borgani2011}. Indeed, this dominance of gravitational effects over baryonic effects was partly the motivation behind the earliest numerical simulations of the non-linear collapse of cosmic structure, such as the work of \citet{press1974}, \citet{white1976} and \citet{gott1979}, who each used $N$-body simulations of collisionless particles to study the build-up of structure, although many early simulations failed to produce adequate amounts of halo substructure. This was attributed to the `over-merging' problem; dark matter subhaloes passing through a larger halo are heavily stripped, and pass below the simulation resolution limit \citep{frenk1996,moore1996a}. Subsequent work, such as that of \citet{moore1998}, was able to resolve this substructure, and confirmed that this over-merging was indeed responsible for the apparent lack of substructure in dark matter-only simulations.

The subsequent development of the $\Lambda$CDM model, plus increases in available computational power, have since allowed for the production of much larger simulations of cosmological volumes. Modern simulations, such as Millennium-XXL \citep{angulo2012}, the Jubilee project \citep{watson2014} and the MultiDark simulations \citep{klypin2016} contain billions of cold dark matter particles in gigaparsec-scale volumes, and allow for detailed studies of the dark matter-dominated formation of large-scale structure.
 
Hydrodynamical simulations build on these ideas by including baryonic material as well as dark matter, in order to model the properties of galaxies, as well as the underlying cosmic structure. These include processes such as gas cooling, star formation and feedback, and employ methods such as smoothed-particle hydrodynamics to better approximate the physical properties of gas \citep{weschler2018}. `Sub-grid physics' is also used to model processes that are below the typical resolution limits of cosmological simulations. Processes such as star-formation and supernova feedback are included in this sub-grid physics, however much of the physics governing these processes is not fully understood. This means that empirical `recipes' must often be included in hydrodynamical simulations, in order to describe this physics (see \citet{springel2010}, \citet{somerville2015} and \citet{vogelsberger2020} for detailed reviews).

The other major drawback of hydrodynamical models is the increased computational time that they require, and so various alternative approaches exist that are less physically motivated, but employ empirical models to save on computing power. These typically utilise $N$-body dark matter-only simulations to simulate the cosmic structure found in hydrodynamical simulations, that are then post-processed to retroactively include the baryonic material. Semi-analytic models are an example of a method that uses $N$-body simulations as its starting point. In these models, baryonic gas is later included in numerically simulated dark matter haloes, and the gas is subsequently evolved using models of gas cooling and star formation to reproduce the evolution of galaxies \citep{benson2001, baugh2006, croton2016}. Other empirical approaches include halo occupation models, in which haloes within a dark matter-only simulation are populated with galaxies statistically, based on a probability distribution that matches galaxies of given properties to corresponding haloes \citep{guo2016, weschler2018}.

These alternatives to hydrodynamical models are generally successful, and recent models have been able to reproduce galaxy properties in impressive detail. For example, the {\sc{GalICS 2.0}} semi-analytic code created by \citet{cattaneo2017} is able to reproduce both the galaxy stellar mass function and Tully-Fisher relation for galaxies in a (100 Mpc)$^{3}$ volume, which had previously been a significant issue for these models \citep{heyl1995, baugh2006}. Similarly, the model of \citet{porter2014} successfully predicts the Fundamental Plane relation for early-type galaxies \citep{djorgovski1987, dressler1987}. The increased speed of both halo occupation models and semi-analytic models over hydrodynamical simulations means that they are particularly useful for exploring large parameter spaces \citep{benson2010, weschler2018}, and can also be used to create huge samples of galaxies, to study large cosmological volumes and generate mock observations \citep{eke2006, frenk2012, somerville2015}. The power of this is demonstrated by \citet{carretero2015}, who have used a halo occupation model and the MICE simulations \citep{crocce2015} to produce mock observations that are being used by the upcoming \textit{Euclid} mission\footnote{The MICE mock galaxy catalogue is publicly available from the CosmoHub database, \url{https://cosmohub.pic.es}.}. Similarly, \citet{knebe2018} have applied three distinct semi-analytic models to the MDPL2 MultiDark simulation \citep{klypin2016}, generating the largest ever public mock galaxy catalogues\footnote{The MultiDark simulations, and the associated mock galaxy catalogues, are publicly available from the CosmoSim database, \url{https://www.cosmosim.org}.}.

The underlying assumption of these models is that the substructure of dark matter-only simulations is valid, compared to a more physical picture involving baryons. However, several studies have indicated regimes in which this may not be the case. Previous work has shown that the cumulative halo mass function is dependent on the baryonic processes that are present in a simulation \citep[see][for example]{cui2012,cui2014}. Similarly, \citet{vandaalen2011} study the matter power spectra of several hydrodynamical simulations from the OWLS project \citep{schaye2010}, as well as the power spectrum of an equivalent dark matter-only simulation. They find that on length scales less than the typical size of clusters ($\lesssim1$ Mpc), the power spectrum amplitude is suppressed in dark matter-only simulations, which is attributed to the cooling and collapsing of baryonic material into dense halo cores in hydrodynamical simulations. This steepened baryonic radial density profile can then cause the dark matter halo to contract, in the same way as was shown by \citet{blumenthal1986}, potentially leading to denser regions of dark matter in hydrodynamical simulations. Other work has found similar results on smaller scales. \citet{jia2020} compare hydrodynamical and dark matter-only simulations of a $10^{14.8}\ h^{-1}M_{\odot}$ halo, at three mass resolutions. They find that the subhalo mass function is steeper in the hydrodynamical simulations, supporting the idea that the substructure in dark matter-only simulations is suppressed on small scales. \citet{libeskind2010} show analogous results, but in even smaller (Local Group-sized) simulations.

\citet{jia2020} also show that subhaloes are more concentrated in their hydrodynamical simulations, again confirming the mechanism of dark matter halo contraction described by \citet{blumenthal1986}. Indeed, other studies have found that halo density profiles are steepened by baryonic effects, leading to haloes being more concentrated in dark matter-only simulations \citep{rasia2004,lin2006}. Additionally, the central regions of dark matter haloes appear to be most strongly affected \citep{cui2014,scaller2015}. Such a mechanism is also supported by the work of \citet{dolag2009}. They show that subhaloes in radiative hydrodynamical simulations, which produce dense stellar regions in the centres of dark matter haloes, are more resistant to the stripping of gas and dark matter than haloes in non-radiative hydrodynamical simulations that lack these stellar cores.

In spite of this, the significance of the effect of baryons on dark matter haloes remains unclear, partly because of its dependence on the models that are implemented \citep{tissera2010,cui2016}. There does exist some disagreement within the literature, with some studies instead finding less substructure in hydrodynamical simulations, which is often attributed to increased tidal disruption in hydrodynamical simulations \citep{zhu2016,garrisonkimmel2017,richings2020}. The effect of baryons at different halo masses is also unclear; \citet{chua2017} show that the subhalo mass function is shallower in hydrodynamical and dark matter-only simulations, in contrast to several other studies, including the recent work of \citet{jia2020}. Other studies show that the presence of baryons simply does not have a strong effect on substructure. In \citet{bahe2019}, the fraction of galaxies being accreted by clusters that survive to redshift $z=0$ is only weakly dependent on whether baryons are included, although they explain that this may be due to the sub-grid physics implemented within their simulations.

In this work, we investigate how cosmic structure and substructure are affected by including baryons in cosmological simulations. We begin by studying galaxies in galaxy clusters, and go on to investigate the outskirts of clusters. We particularly focus on galaxy groups in these outskirts, as a significant fraction of the galaxies that are accreted by clusters join the cluster as members of a galaxy group. \citet{berrier2009} use dark matter-only simulations to find that $30\%$ of galaxies of virial mass\footnote{\citet{berrier2009} use the definition of virial mass laid out by \citet{bryan1998}.} greater than \mbox{$10^{11.5}\ h^{-1}M_{\odot}$} have joined a cluster as part of a group, and $12\%$ as part of a group of six or more galaxies. They also show that these fractions are slightly greater when a lower galaxy mass limit is used, in agreement with subsequent studies \citep{choquechallapa2019}. There is, however, some disagreement in this figure; some studies have found the fraction of infalling galaxies within groups to be as low as $10\%$ \citep{arthur2017}. Others have found that groups can make up almost half of infalling galaxies \citep{mcgee2009}, although it is important to note that this variation is partly down to the way in which groups are defined, which varies between different works. Studies of groups in cluster outskirts are therefore crucial in learning about the growth of clusters, and the histories of galaxies in cluster environments.

Throughout this work, we utilise \threehun\ project, a sample of 324 galaxy clusters taken from a \mbox{$1\ h^{-1}$ Gpc} cosmological volume, and resimulated out to distances of several times the $R_{200}$ of the cluster, where $R_{200}$ is the radius within which the mean density of a cluster is equal to 200 times the critical density of the Universe. We first examine the radial number density profiles of galaxies in galaxy clusters, and of galaxy groups in the cluster outskirts. These clusters have all been simulated with both full hydrodynamics, and using only cold dark matter. We then look at the phase space of galaxies in groups that are on their first infall into a cluster, to study how this distribution is affected by the inclusion of baryonic material in simulations. 

This paper is structured as follows: in \Sec{sec:methods} we introduce the simulation data and analysis used in this work. In \Sec{sec:results} we present our results, and in \Sec{sec:discussion} we discuss some of the causes and effects of the differences we find between the two types of simulations. Finally, we summarise our findings in \Sec{sec:conclusions}.

\section{Simulations \& Numerical methods}
\label{sec:methods}

\subsection{Simulation data}
\label{sec:hydro}

In this work we use data from \threehun\ project, a suite of 324 galaxy clusters, forming a mass-complete sample taken from the dark-matter-only MDPL2 MultiDark simulation \citep{klypin2016}. This simulation uses \textit{Planck} cosmology (\mbox{$\Omega_{\rm{M}}=0.307$}, \mbox{$\Omega_{\rm{B}}=0.048$}, \mbox{$\Omega_{\Lambda}=0.693$}, \mbox{$h=0.678$}, \mbox{$\sigma_{8}=0.823$}, \mbox{$n_{\rm{s}}=0.96$}) \citep{planck2016} to simulate a box with sides of comoving length \mbox{$1\ h^{-1}$ Gpc}. To generate the data for \threehun\ project, the 324 most massive clusters at $z=0$ were chosen, and all particles within \mbox{$15\ h^{-1}$ Mpc} \mbox{($\sim10R_{200}$)} of each cluster's centre at \mbox{$z=0$} were traced back to their initial positions. Each cluster was then resimulated from its initial conditions with full baryonic physics, by splitting each dark matter particle into a dark matter and a gas particle, with masses set by the baryonic matter fraction of the Universe. Lower-resolution particles were used beyond \mbox{$15\ h^{-1}$ Mpc} to model any effects of large-scale structure. The simulations were carried out using the {\sc{gadgetX}} code, a modified version of the {\sc{gadget-3}} code, which uses a smoothed-particle hydrodynamics scheme to fully evolve the gas component of the simulations \citep{springel2005a, beck2016}. 

Additionally, the 324 clusters were also resimulated using only dark matter. The same dark matter particle masses were used as in the original MDPL2 simulation, but the simulations run using the {\sc{gadgetX}} code, as opposed to {\sc{gadget-2}}, which was used in MDPL2 \citep{klypin2016}. These clusters were hence evolved from the same initial conditions as their hydrodynamical counterparts, and using the same simulation code and analysis. This allows us to make like-for-like comparisons between two simulations of the same clusters, showing the effects of baryonic physics on the dynamics of clusters. Finally, we simulated one of the clusters from \threehun\ sample two further times, in order to investigate resolution effects. We hereafter refer to this cluster by its ID, \mbox{`{\sc{cluster}}\_0002'}. This cluster has also been simulated hydrodynamically with a shorter gravitational softening length for stellar particles, and in dark matter-only with a factor of eight increase in resolution. These additional simulations are detailed in \Sec{sec:resolution}, and a summary of the particle data is given in \Tab{tab:partdata}.

\begin{table*}
	\centering
	\caption{Parameters for the four classes of simulations used in this work. Columns 2-5 represent the mass of dark matter particles, $m_{\rm{DM}}$, and of gas particles, $m_{\rm{gas}}$, in the central high-resolution region, the Plummer equivalent gravitational softening length for both gas and dark matter particles, $\epsilon_{\rm{DM,gas}}$, and the  Plummer equivalent gravitational softening length for star particles, $\epsilon_{\rm{stars}}$. The bottom two rows only apply to \mbox{`{\sc{cluster}}\_0002'}, which was used to test resolution effects.}
	\label{tab:partdata}
	\begin{tabular}{ccccc} 
		\hline
		Simulation & $m_{\rm{DM}}\ /\ 10^{8}h^{-1}M_{\odot}$ & $m_{\rm{gas}}\ /\ 10^{8}h^{-1}M_{\odot}$ & $\epsilon_{\rm{DM,gas}}$ / $h^{-1}$ kpc & $\epsilon_{\rm{stars}}$ / $h^{-1}$ kpc\\
		\hline
		Hydrodynamical & 12.7 & 2.36 & 6.5 & 5\\
		Dark matter-only & 15 & -- & 6.5 & --\\
		Reduced softening & 12.7 & 2.36 & 6.5 & 1\\
		High-res DM-only & 1.88 & -- & 3.25 & --\\
		\hline
	\end{tabular}
\end{table*}

The final dataset of 324 clusters range in mass from \mbox{$M_{200}=5\times10^{14}\ h^{-1}M_{\odot}$} to \mbox{$M_{200}=2.6\times10^{15}\ h^{-1}M_{\odot}$}, where $M_{200}$ is the mass contained within a sphere of radius $R_{200}$. The hydrodynamical clusters consist of dark matter and gas particles, and star particles of variable masses, typically with \mbox{$m_{\rm{star}}\sim4\times10^{7}\ h^{-1}M_{\odot}$}, produced by the stochastic star-formation that is implemented by {\sc{gadgetX}} \citep{tornatore2007,murante2010,rasia2015}. \threehun\ dataset is described in more extensive detail in \citet{cui2018}, and has been used in previous studies to examine cluster density profiles \citep{mostoghiu2019,li2020}, environments \citep{wang2018,kuchner2020}, ram pressure \citep{arthur2019}, the hydrostatic equilibrium mass bias \citep{ansarifard2020}, backsplash galaxies \citep{haggar2020}, and the shapes and alignments of galaxies \citep{knebe2020}.

\subsubsection{Galaxy identification and tree-building}
\label{sec:tree}

The halo merger trees used in this work were produced using the same methods as detailed in \citet{haggar2020}. For each cluster, 129 snapshots were saved between \mbox{$z=16.98$} and \mbox{$z=0$}. These were then processed using the {\sc ahf}\footnote{\url{http://popia.ft.uam.es/AHF}} halo finder, to detect the haloes and subhaloes present in each snapshot (see \citet{gill2004} and \citet{knollmann2009} for further details on {\sc ahf}). The version of {\sc ahf} used in this work accounts for gas, stars and dark matter, and returns the position and velocity of each halo and subhalo, as well as properties such as their radii, and their masses in gas, stars and dark matter.

The halo merger trees were built using {\sc mergertree}, a tree-builder designed as part of the {\sc ahf} package. For each halo in a given snapshot, {\sc mergertree} identifies a main progenitor, plus other progenitors, by looking at haloes in all previous snapshots that share particles with it. By considering all prior snapshots when searching for a progenitor (rather than just the previous one), this tree-builder is able to `patch' over gaps in the tree that would otherwise result in halo branches being truncated. This property is particularly useful for studying groups and cluster substructure, as small subhaloes passing through the dense centre of a larger halo are more challenging to identify \citep{onions2012}. We also place a factor of two limit on the change in dark matter mass between snapshots, such that no halo can more than double in dark matter mass between successive snapshots. This helps to prevent non-physical `mismatches' that can be caused by a subhalo located close to the centre of a larger halo (as discussed in \Sec{sec:subsample}). Further details of {\sc ahf} and {\sc mergertree} can be found in \citet{knebe2011b} and \citet{srisawat2013}.

\subsection{Sub-sample of clusters}
\label{sec:subsample}

In the first part of this work, \Sec{sec:radial_dens}, we use the full set of 324 cluster simulations at $z=0$. However, we identify a sample of the hydrodynamical simulations for which we do not have good galaxy tracking data, and so are not suitable for analysis at $z>0$. Consequently, we choose to omit these from our analysis in \Sec{sec:phase_space}, which uses simulation data from before $z=0$. This cluster sample is the same as in \citet{haggar2020}, work that also uses simulation data from $z>0$.

The {\sc mergertree} code uses a merit function ($M_{\rm{i}}$, described in Table B1 of \citet{knebe2013}) to find the most likely progenitors of haloes in preceding snapshots, and build the trees. This merit function is used to `patch' over the gaps in the merger tree, but some halo links between snapshots can be assigned incorrectly, which can lead to an apparent `jump' in position of the cluster halo. Although such events are uncommon and typically only affect a small number of snapshots, they can be a major issue when tracking the times and positions of groups entering a cluster. These `jumps' can also occur at late times (after \mbox{$z=1$}), which is particularly problematic in this work, as most of the galaxy infalls take place at late times; for clusters that can be tracked back to \mbox{$z=5$}, approximately $80\%$ of infall events occur after \mbox{$z=1$}. 

These merger tree mismatches are particularly common during a major merger of the main cluster halo. This is described by \citet{behroozi2015}, who show that two merging haloes of similar size can be accidentally switched by a tree-builder, leading to a `flip-flopping' effect, in which the sizes and positions of haloes appear to change suddenly and dramatically. Of our sample of 324 hydrodynamical clusters, we find 59 whose position (in box coordinates) changes by $>0.5R_{200}(z)$ between two snapshots after $z=1$. We find that this distance is non-physical, given the typical time elapsing between snapshots at this redshift in our simulations \mbox{($\sim0.3$ Gyr)}. 

We also find 17 hydrodynamical clusters whose main branch on the merger tree (that is, the evolution of the main cluster halo) cannot be tracked back to further than $z=0.5$. These cases are largely due to a missing link in the merger tree, resulting in the history of the cluster before this link being lost. Although they still contain some groups that have fallen in at late times, we choose to also remove these clusters from our analysis, as they do not give a good, unbiased sample of groups that have entered a cluster throughout its whole history. Nine of these clusters also experience large jumps in their position (as discussed above), resulting in a total of 67 clusters for which we have relatively poor tracking data. 

Our remaining 257 hydrodynamical clusters have $M_{200}$ masses (dark matter, gas and stars, including subhaloes) between \mbox{$5\times 10^{14}\ h^{-1}M_{\odot}$} and \mbox{$2.6\times 10^{15}\ h^{-1}M_{\odot}$}, with a median value of \mbox{$8\times 10^{14}\ h^{-1}M_{\odot}$}. The radii ($R_{200}$) of these clusters range from \mbox{$1.3\ h^{-1}$ Mpc} to \mbox{$2.3\ h^{-1}$ Mpc}, with a median of \mbox{$1.5\ h^{-1}$ Mpc}. For consistency, this same sample of 257 dark matter-only clusters are also used in \Sec{sec:phase_space}.

\subsection{Galaxy and group selection}
\label{sec:galgroups}

In this work, the word `galaxy' refers to all the components of an object in the hydrodynamical simulations, including its stellar and dark matter components. These can either be individual objects, or may be bound to a group. We use the word `galaxy' to describe these in a general context, but in the specific context of our dark matter-only simulations, we hereafter refer to these objects as `subhaloes' instead.

Throughout this work, we place a limit on the total (dark matter, gas and stars) mass of galaxies/subhaloes within the simulations of \mbox{$M_{200}\geq10^{10.5}\ h^{-1}M_{\odot}$}. This corresponds to approximately $100$ particles in the \mbox{$15\ h^{-1}$ Mpc} high-resolution region surrounding each cluster. Using a mass cut (rather than a particle number cut) removes any bias towards the hydrodynamical simulations, as the gas/star particles have lower masses than the dark matter particles, and so lower mass objects can be found in these simulations. We also remove all objects from the hydrodynamical simulations that contain more than $30\%$ of their mass in stars. These haloes are typically found very close to the centre of a larger halo, meaning that much of their dark matter has been stripped (evidence of this tidal stripping in \threehun\ simulations is presented in \citet{knebe2020}). The remnants of this process are very compact objects with high stellar mass fractions, whose properties (such as their radii and masses) are not well-determined by our halo finder. Given this, and the fact that these haloes make up only $1\%$ of all haloes within $5R_{200}$ of the clusters, we make the decision to remove these objects from our analysis. By taking this approach, and not including an absolute stellar mass limit in the hydrodynamical simulations, we keep all objects with a total mass above \mbox{$10^{10.5}\ h^{-1}M_{\odot}$}, and therefore ensure that the hydrodynamical and dark matter-only simulations are equivalent.

\subsubsection{Group identification}
\label{sec:group_id}

In this work, we identify galaxy groups in the simulations by considering each galaxy, and determining how many other galaxies in the same snapshot are associated with it. If the galaxy has four or more other galaxies associated with it, we take it to be the host halo of a group. Galaxies are defined as being associated with this `group host' (and hence a member of the group) if they satisfy the same criteria as \citet{han2018} and \citet{choquechallapa2019} use to define galaxy groups; the total (dark matter, gas, and stars) mass of a galaxy must be less than that of its group host, and the galaxy must satisfy the criterion below:

\begin{equation}
    \frac{v^2}{2}+\Phi\left(r\right)<\Phi\left(2.5R_{200}^{\rm{grp,h}}\right)\,.
    \label{eq:bounded}
\end{equation}

Here, $\Phi(r)$ represents the gravitational potential due to the group host at a distance $r$ from its centre, and $v$ is the relative velocity of a galaxy with respect to its group host. $R_{200}^{\rm{grp,h}}$ is the radius of the group host halo (this is distinct from the radius of the host cluster that is present in each of the simulations, which is subsequently referred to by $R_{200}^{\rm{clus,h}}$). We take any galaxies that are less massive than their group host and that satisfy this criterion to be bound members of this group. 

This criterion means that galaxies bound to their group host can be found as far as $2.5R_{200}^{\rm{grp,h}}$ from the centre of the group, providing their velocity relative to the group is sufficiently small. Similarly, fast-moving galaxies near to the pericentre of their orbit in the group are also included. However, the presence of this velocity-dependent term means that `fly-by' galaxies, which pass near to the group but are not bound to it, are excluded from this selection. These galaxies are equivalent to the `renegade subhaloes' identified by \citet{knebe2011c} in simulations of the Local Group.

\subsubsection{Infalling groups}
\label{sec:infalling}

Throughout this work, we consider galaxy groups in two regimes. In \Sec{sec:radial_dens}, we identify galaxy groups located in the region we refer to as the `cluster outskirts', between \mbox{$[R_{200}^{\rm{clus,h}}, 5R_{200}^{\rm{clus,h}}]$} from the cluster centre at $z=0$. Then, in \Sec{sec:phase_space}, we instead identify infalling galaxy groups, at all redshifts, again using the same methods as \citet{han2018} and \citet{choquechallapa2019}. Using the halo merger trees, we first identify all objects that have just fallen into the cluster, taken from the whole history of the cluster; that is, we find galaxies that were at a distance greater than $R_{200}^{\rm{clus,h}}$ from the cluster centre in one snapshot, but are within $R_{200}^{\rm{clus,h}}$ in the following snapshot. We refer to these objects as the `infalling' galaxies. 

\begin{figure}
	\includegraphics[width=\columnwidth]{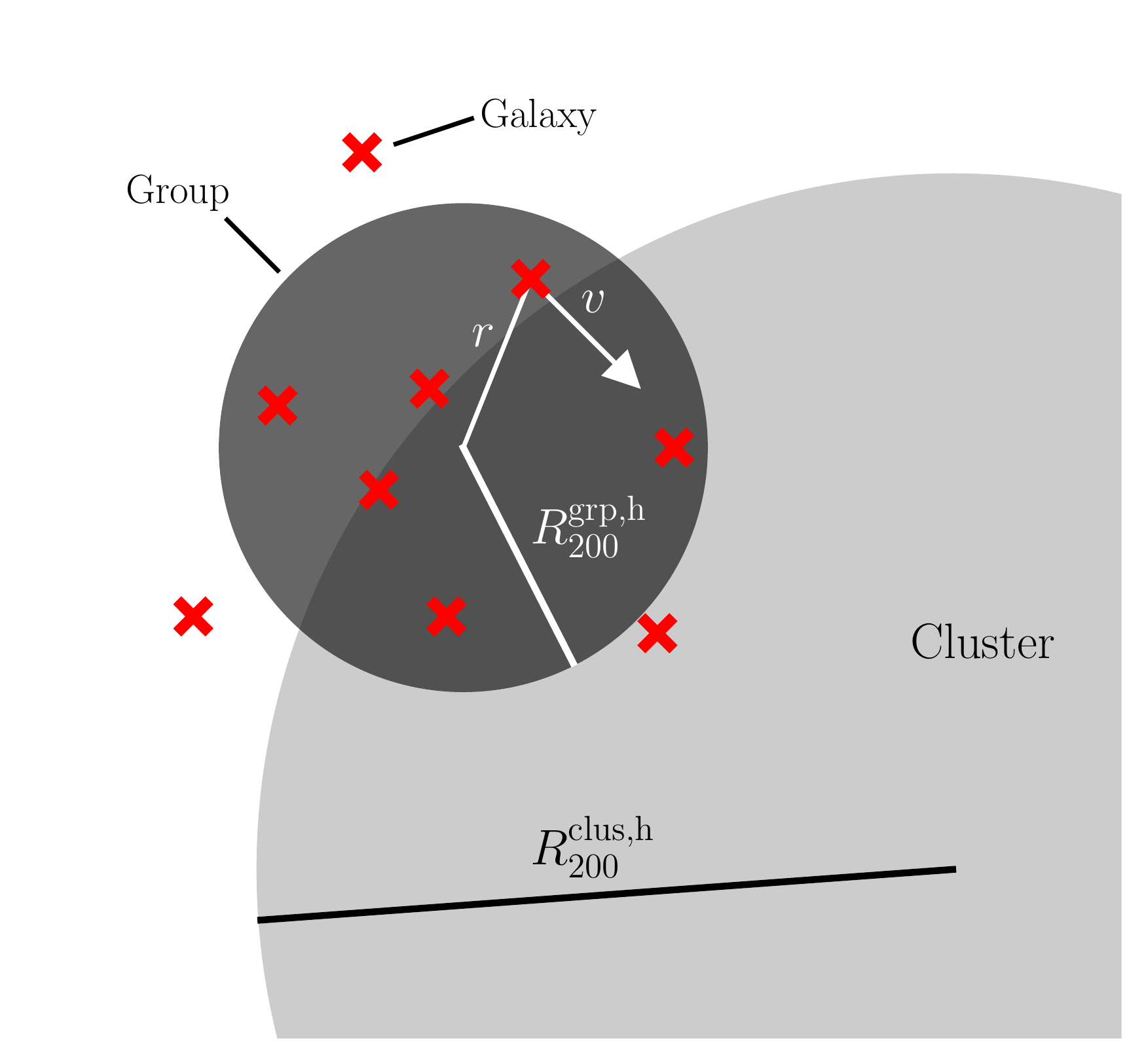}
    \caption{Schematic of a galaxy group halo (dark circle) falling into a cluster (light circle). The centre of the host group has just passed within $R_{200}^{\rm{clus,h}}$ of the cluster. Red crosses represent galaxies that are members of this group; note that these can be outside of the cluster at the time of group infall, and that they can be outside of $R_{200}^{\rm{grp,h}}$ based on the definition in \Eq{eq:bounded}. The position, $r$, and velocity, $v$, of one galaxy relative to its host group are also labelled. It is this configuration that is used in the second part of our analysis (\Sec{sec:phase_space}).}
    \label{fig:schematic}
\end{figure}

For each infalling galaxy, we consider the first snapshot in which it has passed within $R_{200}^{\rm{clus,h}}$, and use the method in \Sec{sec:group_id} to determine if any galaxies are bound to it at this time, thereby making the infalling galaxy a group host halo. This galaxy, and any galaxies that are bound to it, then make up the infalling group. A schematic is given in \Fig{fig:schematic}, showing the configuration of an infalling group. Haloes that are on a second (or subsequent) infall are excluded, so that our sample of groups consists only of those entering the cluster for the first time. These repeat infallers make up only $13\%$ of the infalling galaxies, and less than $1\%$ of the bound groups that we identify.

\section{Results}
\label{sec:results}

\subsection{Radial density profiles of clusters and groups}
\label{sec:radial_dens}

\begin{figure}
	\includegraphics[width=\columnwidth]{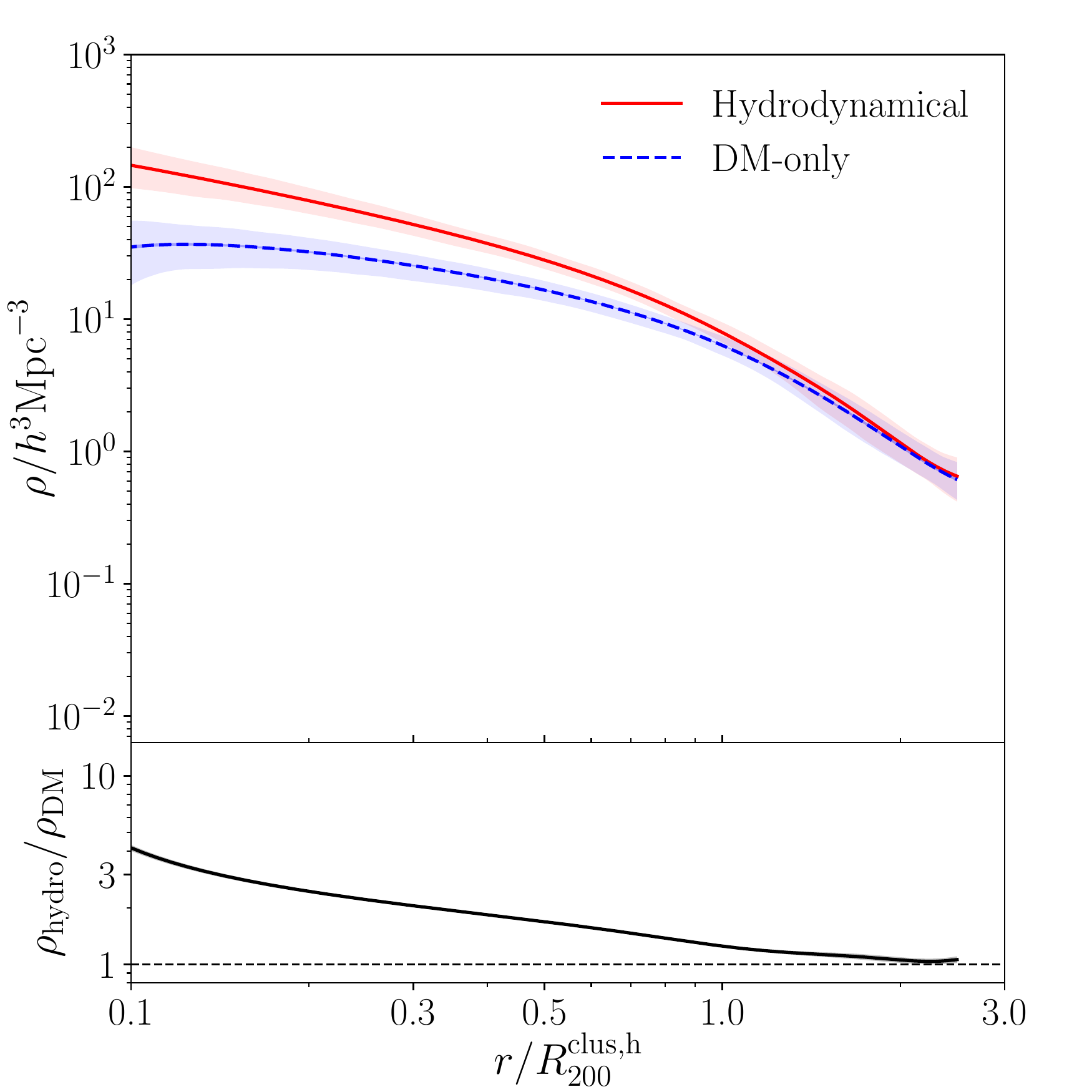}
    \caption{Radial number density, $\rho$, of galaxies gravitationally bound to clusters, for 324 hydrodynamical clusters and dark matter-only clusters at $z=0$ (top panel). Light shaded regions show the $1\sigma$ spread between clusters, and dark shaded regions represent $1\sigma$ uncertainty in the mean radial number density profile, although these are mostly too small to be seen. Bottom panel shows fractional residuals, which we define in the main text.}
    \label{fig:dm_hydro_cluster}
\end{figure}

\Fig{fig:dm_hydro_cluster} shows the number density profile of galaxies that are bound (according to the criterion in \Eq{eq:bounded}) to a host cluster. For consistency, we use the same criteria to define galaxies bound to the cluster as we use for defining group member galaxies, but we note that this bound population of galaxies represents almost all galaxies in the cluster; over $99\%$ of galaxies within $R_{200}$ of a cluster are gravitationally bound. Note also that the distances from the cluster centre, $r$, are given in units of the cluster radii $R_{200}^{\rm{clus,h}}$ and so are normalised between clusters, but the number densities are given in units of \mbox{$h^{3}\ \rm{Mpc}^{-3}$}. For each cluster we generate a kernel density estimation (KDE) of the distribution of galaxies. Using a KDE with an optimised bandwidth provides a smoothed distribution, and removes most of the effects of bin selection that can impact a histogram. We then average the KDEs across all clusters. As this analysis only requires data from $z=0$, we use data from all 324 clusters.

Clusters in the dark matter-only simulations have a deficit of subhaloes in their central regions, relative to the hydrodynamical cluster simulations. The deficit decreases with increasing distance from the cluster centres, and the two profiles are indistinguishable outside of \mbox{$2.1R_{200}^{\rm{clus,h}}$}, as shown by the residual in the bottom panel of \Fig{fig:dm_hydro_cluster}. This is equal to the ratio of the galaxy number density profile in the hydrodynamical clusters (given by $\rho_{\rm{hydro}}$) to the density profile in the dark matter-only simulations ($\rho_{\rm{DM}}$). Consequently, in regions where this residual is equal to one, the number density of galaxies is equal in the two types of simulation. The error bounds on this residual come from the uncertainty in the mean density profiles (the dark shading in the top panel). This residual demonstrates that the substructure of hydrodynamical simulations is only reproduced in dark matter-only simulations in the outer regions of clusters.

Some of the difference between the number densities of galaxies in the cluster outskirts can be explained by the inclusion of backsplash galaxies, which have previously passed within $R_{200}^{\rm{clus,h}}$ of a cluster, but now exist beyond this radius in the cluster outskirts. If we exclude these backsplash galaxies from our analysis, we instead find that the number density profiles of the hydrodynamical and dark matter-only clusters agree at radii beyond $1.4R_{200}^{\rm{clus,h}}$. This indicates that backsplash galaxies are less likely to survive the passage through a cluster in dark matter-only simulations, as they contribute less to the number density of galaxies in these simulations. Indeed, we find that in the dark matter-only simulations, an average of $45\%$ of galaxies in the radial region \mbox{$[R_{200}^{\rm{clus,h}}, 2R_{200}^{\rm{clus,h}}]$} are backsplash galaxies, compared to $51\%$ in the hydrodynamical simulations \citep[see also][]{haggar2020}\footnote{\citet{haggar2020} gives a slightly different value for the average backsplash fraction in hydrodynamical simulations, as that work includes a galaxy stellar mass cut.}. As this difference in backsplash fraction is small, it can only explain a small part of the number density deficit in dark matter-only simulations. At cluster distances greater than $1.4R_{200}^{\rm{clus,h}}$ the difference in number density between the two simulations is less than $15\%$, even when including backsplash galaxies. Within this radius, the increased deficit of galaxies in the dark matter-only simulations cannot be fully explained by these missing backsplash galaxies.

Unlike the mass density profiles of these clusters \citep{mostoghiu2019}, the number density profiles in \Fig{fig:dm_hydro_cluster} are not well-described by an NFW profile \citep{navarro1996}, particularly in the cluster outskirts. This is potentially because of the boundness criteria we employ, which place strict limits on the velocities of galaxies in the outer regions of the clusters. However, in the radial region \mbox{$[0.2, 2.0]R_{200}^{\rm{clus,h}}$}, the cluster number density profiles are well-described by an Einasto profile \citep{einasto1965,navarro2004}. Specifically, we use the form of \citet{springel2008},

\begin{equation}
    \rho\left(r\right)=\rho_{0}\,{\rm{exp}}\left(-\frac{2}{\alpha}\left[\left(\frac{r}{r_{0}}\right)^{\alpha}-1\right]\right)\,,
    \label{eq:einasto}
\end{equation}
where $\rho_{0}$, $r_{0}$ and $\alpha$ are free parameters describing the profile. The values of these parameters for the radial density profiles of clusters in both the hydrodynamical and dark matter-only simulations are given in \Tab{tab:einasto}. The parameter $\alpha$ describes the curvature of the profile, and its greater value for the dark matter-only clusters demonstrates how this profile is shallower near to the centre of a cluster, but drops off equally steeply at greater radii.

\begin{table*}
	\centering
	\caption{Parameters for the best-fit Einasto profiles (\Eq{eq:einasto}), for the hydrodynamical and dark matter-only simulations of the number density profiles of galaxy clusters (\Fig{fig:dm_hydro_cluster}) and galaxy groups in the cluster outskirts (\Fig{fig:dm_hydro_all_groups}). For clarity, we have not included these fits in the relevant figures.}
	\label{tab:einasto}
	\begin{tabular}{ccccc} 
		\hline
		 & Simulation & $\rho_{0}\ /\ h^{3}$ Mpc$^{-3}$ & $r_{0}$ / $R_{200}^{\rm{h}}$ & $\alpha$\\
		\hline
		\multirow{2}{*}{Clusters} & Hydrodynamical & $11.709\pm0.006$ & $0.8339\pm0.0003$ & $0.640\pm0.001$\\
		& DM-only & $5.479\pm0.003$ & $1.0747\pm0.0004$ & $0.878\pm0.001$ \\
		\multirow{2}{*}{Groups} & Hydrodynamical & $33.43\pm0.04$ & $0.5303\pm0.0002$ & $0.931\pm0.001$\\
		& DM-only & $7.46\pm0.01$ & $0.7615\pm0.0003$ & $1.302\pm0.002$ \\
		\hline
	\end{tabular}
\end{table*}

\begin{figure}
	\includegraphics[width=\columnwidth]{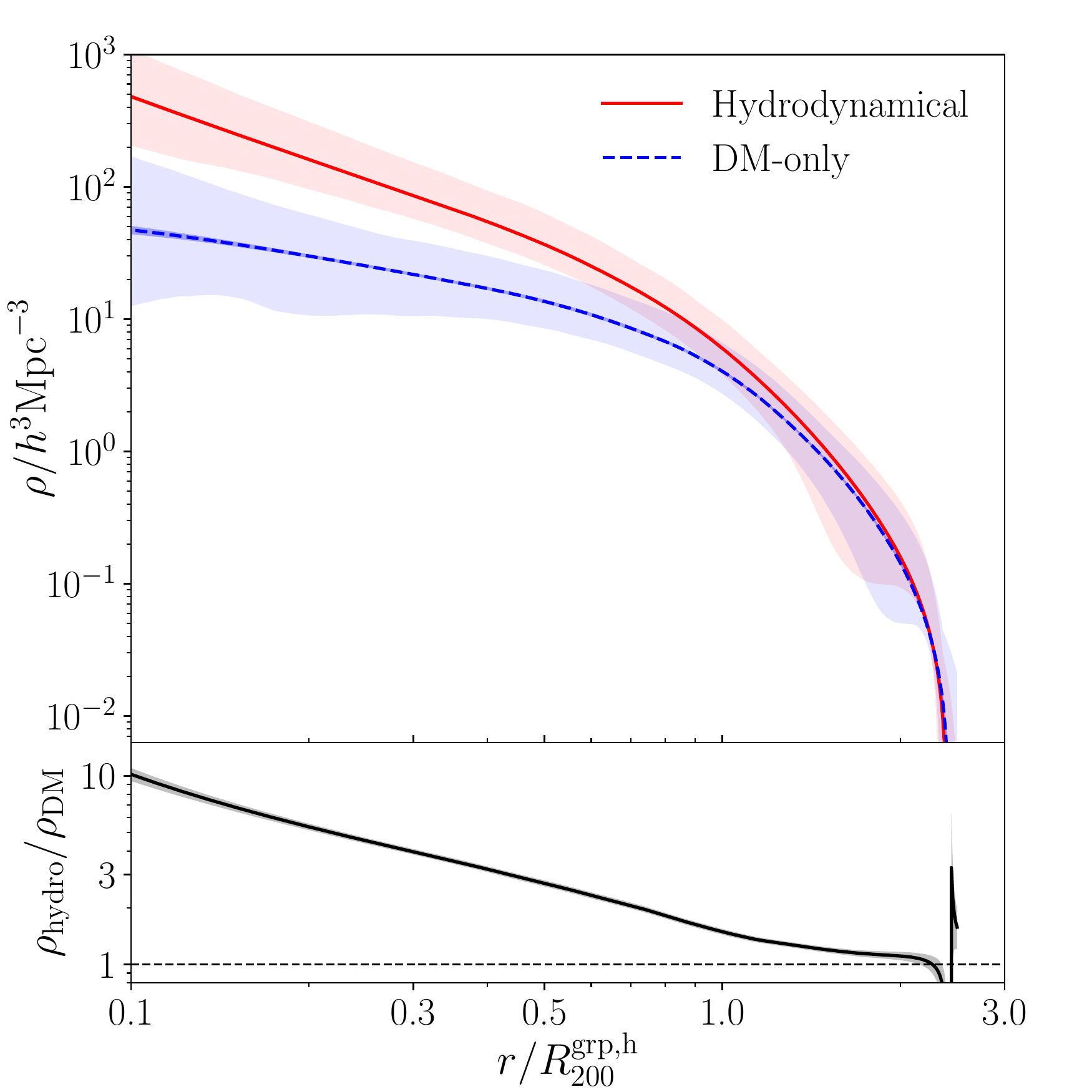}
    \caption{Radial number density, $\rho$, of galaxies in groups located in cluster outskirts at $z=0$ (top panel). Includes all groups in the radial range \mbox{$[R_{200}^{\rm{clus,h}}, 5R_{200}^{\rm{clus,h}}]$}, consisting of between five and 50 galaxies. Light shaded regions show the $1\sigma$ spread between groups, and dark shaded regions represent $1\sigma$ uncertainty in the mean radial number density profile, although these are mostly too small to be seen. Bottom panel shows fractional residuals.}
    \label{fig:dm_hydro_all_groups}
\end{figure}

\Fig{fig:dm_hydro_all_groups} shows data equivalent to that in \Fig{fig:dm_hydro_cluster}, except that instead of the density profiles of the clusters, it gives the mean radial number density profile of galaxy groups located in the cluster outskirts, between \mbox{$[R_{200}^{\rm{clus,h}}, 5R_{200}^{\rm{clus,h}}]$} from the cluster centre, at $z=0$. Groups that have between five and 50 galaxies (including the group host galaxy) that each satisfy our mass criteria are included. Here, we generate a KDE for each individual galaxy group, and average these across all groups in the whole sample of 324 clusters. Throughout our analysis of the `group members', we exclude the `host galaxy' at the centre of each group's host halo.

As is also the case for the clusters in \Fig{fig:dm_hydro_cluster}, groups produced in the dark matter-only simulations have a deficit of subhaloes, relative to groups in the hydrodynamical simulations. This deficit is strongest in the central regions, and as shown by the residuals in the bottom panel in \Fig{fig:dm_hydro_all_groups}, the two profiles agree within uncertainties in the group outskirts, beyond \mbox{$2.1R_{200}^{\rm{grp,h}}$}. The apparent spike in the residual at \mbox{$r\approx 2.5R_{200}^{\rm{grp,h}}$} is due to small number statistics, as there are very few galaxies at this distance from the group centres. In the hydrodynamical simulations, we find a very small dependence of the number density profile on the distance of the groups from the cluster centre -- the number density of galaxies is approximately $20\%$ lower in groups within \mbox{$2R_{200}^{\rm{clus,h}}$} of the cluster centre, compared to those beyond this distance. We find no significant systematic dependence on the cluster distance in the dark matter-only simulations.
The shapes of the radial number density profiles for both the hydrodynamical and dark matter-only groups are different to those of the profiles of the clusters. Generally the cluster profiles are flatter, with a much shallower decrease in number density beyond $R_{200}$. However, in the radial region \mbox{$[0.3, 2.0]R_{200}^{\rm{grp,h}}$}, the group number density profiles can also be well-described by an Einasto profile -- the parameters of the best-fit profiles are given in \Tab{tab:einasto}. As is also the case for the cluster profiles in \Fig{fig:dm_hydro_cluster}, the change in the slope is sharper in the dark matter-only groups.

\subsubsection{Masses and radii of groups}
\label{sec:masses_radii}

We have also investigated whether the radial number density profiles of groups in the cluster outskirts are dependent on the mass of the group host haloes, $M_{200}^{\rm{grp,h}}$. The median mass of the host halo for groups in the cluster outskirts is \mbox{$10^{13.5\pm0.4}\ h^{-1}M_{\odot}$} in the dark matter-only simulations, and \mbox{$10^{13.2\pm0.4}\ h^{-1}M_{\odot}$} in the hydrodynamical simulations (these error bars represent the $1\sigma$ spread of the data). The range of $M_{200}^{\rm{grp,h}}$ is approximately two orders of magnitude; the range of group masses is \mbox{$[10^{12.3},10^{14.5}]\ h^{-1}M_{\odot}$} in the dark matter-only simulations, and \mbox{$[10^{12.0},10^{14.2}]\ h^{-1}M_{\odot}$} in the hydrodynamical simulations, for the groups that we have selected (with between five and 50 members). We generally find that the number density of galaxies is less in larger groups, and that the number density profiles are flatter. In large groups the radial number density profile becomes closer to the cluster profile, particularly in the inner regions of the group. Splitting the galaxy groups into three categories based on halo mass, we find that the variation in radial number density between the most massive and least massive groups is relatively small, typically less than a factor of three.

Although the average mass of group haloes is slightly greater in our dark matter-only simulations, this is not enough to fully explain the significantly flatter radial density profile in these simulations. The difference in average group mass between our hydrodynamical and dark matter-only simulations is 0.3 dex, which is small compared to the variation in group mass within each simulation (approximately 2 dex). It is therefore not enough to account for the difference in number density between the simulations, which is a factor of 10 in the central regions of the groups. We discuss the dependence of the radial number density on the group halo mass in more detail in \App{sec:appendix_grp_mass}.

\begin{figure}
	\includegraphics[width=\columnwidth]{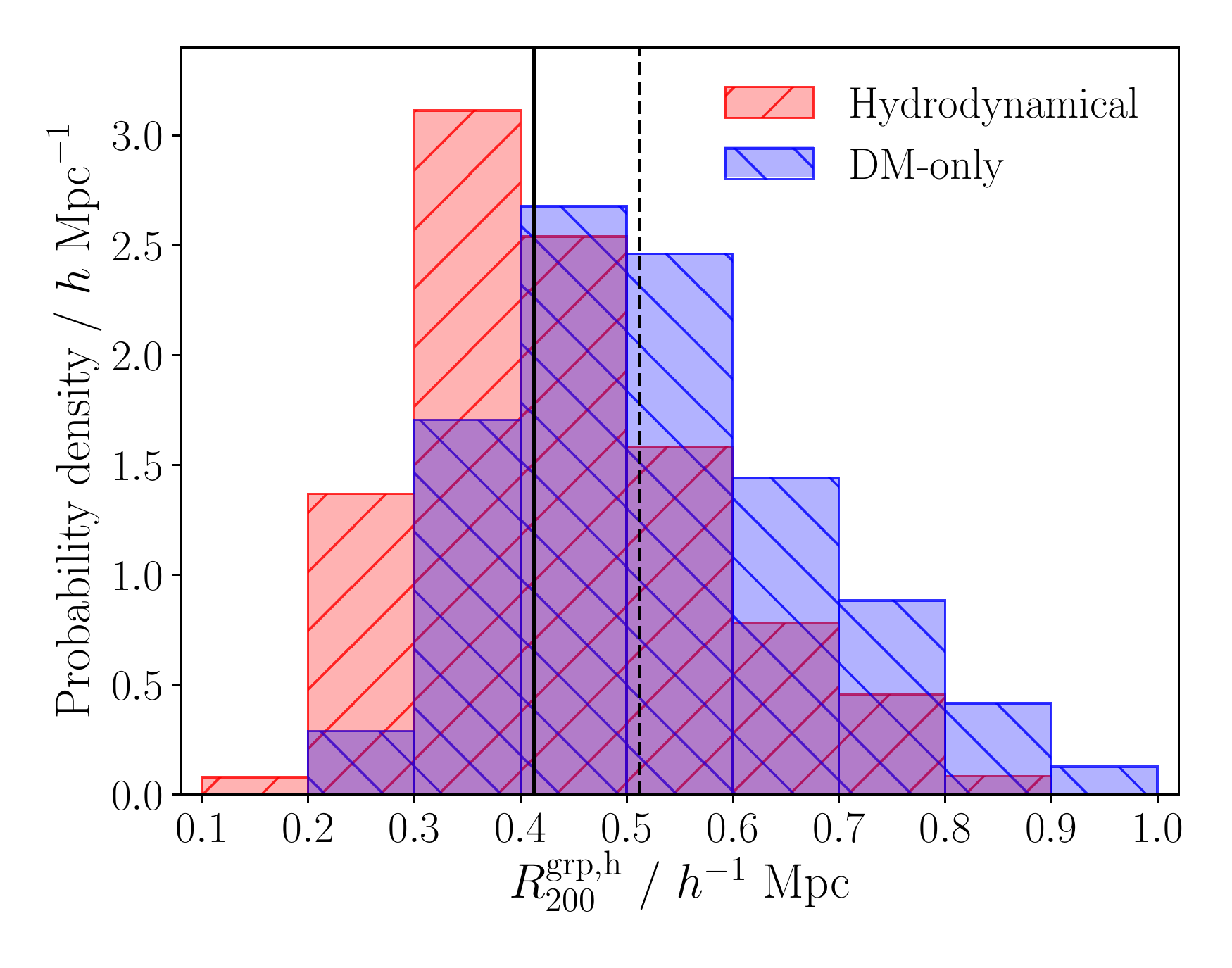}
    \caption{Histogram of radii, $R_{200}^{\rm{grp,h}}$, of groups with between five and 50 members, located between $R_{200}^{\rm{clus,h}}$ and \mbox{$5R_{200}^{\rm{clus,h}}$} from a cluster, in hydrodynamical and dark matter-only simulations. Solid and dashed vertical lines show the median group radius in the hydrodynamical and dark matter-only simulations, respectively.}
    \label{fig:r200_hist}
\end{figure}

Naturally, given that groups in the dark matter-only simulations have slightly greater masses, we also find that these groups have greater average radii than in the hydrodynamical simulations. This is shown in \Fig{fig:r200_hist}; the median radius, $R_{200}^{\rm{grp,h}}$, for groups of between five and 50 members is \mbox{$0.51^{+0.17}_{-0.13}\ h^{-1}$ Mpc} in the outskirts of the dark matter-only clusters, compared to \mbox{$0.41^{+0.17}_{-0.10}\ h^{-1}$ Mpc} in the hydrodynamical simulations (these error bars also represent the $1\sigma$ spread of the data). This difference in median radius is equivalent to a $92\%$ increase in the median volume of these groups, meaning that although the number density of galaxies in the group outskirts is the same in both data sets, this would result in a greater number of galaxies in the outskirts of dark matter-only groups.
Indeed, we find that the median number of group members outside of $R_{200}^{\rm{grp,h}}$ is \mbox{$2^{+4}_{-2}$} in the hydrodynamical simulations, and \mbox{$3^{+4}_{-2}$} in the dark matter-only, demonstrating this increase, although it is much smaller than the spread in the data. The median total number of group members is the same in the hydrodynamical and dark matter-only simulations (\mbox{$8^{+11}_{-3}$} and \mbox{$8^{+9}_{-3}$} respectively). The difference in the average group halo radius is not seen in the radii of the host clusters, $R_{200}^{\rm{clus,h}}$; on average, there is less than a $1\%$ variation in the radius of each cluster between the hydrodynamical and dark matter-only simulations. 

\Fig{fig:r200_hist} also shows that some of the groups, particularly in the hydrodynamical simulations, have very small radii \mbox{($<200\ h^{-1}$ kpc)}. This is due to the fact that we do not apply a lower mass limit to the group host haloes (besides the $10^{10.5}\ h^{-1}M_{\odot}$ mass limit that is applied to all haloes). Despite this, even the smallest groups that we identify have \mbox{$R_{200}^{\rm{grp,h}}\approx160\ h^{-1}$ kpc}, \mbox{$M_{200}^{\rm{grp,h}}\approx10^{12}\ h^{-1}M_{\odot}$}, and typically contain approximately five galaxies, which we consider large enough to still represent physical galaxy groups. In fact, certain galaxy groups, such as Hickson Compact Groups \citep{hickson1982}, can contain similar numbers of galaxies to this within an even smaller radius, sometimes less than \mbox{50 kpc} \citep{barton1996}.

\subsection{Phase space of infalling groups}
\label{sec:phase_space}

The results from \Fig{fig:dm_hydro_cluster} show that the inclusion of baryonic material affects the substructure in galaxy clusters, and \Fig{fig:dm_hydro_all_groups} shows that this effect is even stronger in galaxy groups located in the cluster outskirts. As described in \Sec{sec:intro}, a significant fraction of galaxies within clusters have previously been members of a group, that has since been accreted by a cluster \citep{berrier2009, mcgee2009, arthur2017}. Indeed, in our hydrodynamical simulations we find that over the history of a cluster, an average of \mbox{$(14.2\pm0.2)\%$} of galaxies that enter $R_{200}^{\rm{clus,h}}$ do so as part of a bound group (this error represents the uncertainty in the mean; throughout most of this work we instead quote the spread in the data). For the dark matter-only clusters, \mbox{$(6.2\pm0.1)\%$} of subhaloes are accreted as members of groups; this lower fraction is expected, given the lower number density of group members in the dark matter-only simulations, shown by \Fig{fig:dm_hydro_all_groups}. The fraction of galaxies accreted in hydrodynamical groups is similar to other work that uses similar sized hydrodynamical clusters \citep{arthur2017}, and is in line with the typical mass fraction found in subhaloes \citep{gao2011}. The greater fraction that is found in some other work \citep[e.g.][]{mcgee2009} is likely caused by their use of different mass limits for galaxies and the group host (note that we use the same limit for these).

The accretion of galaxy groups is therefore an important part of the growth of galaxy clusters, and this has motivated a wide range of studies into the properties of these groups. For example, \citet{choquechallapa2019} use dark matter-only simulations to identify groups using the same method as this work, and then look at the phase space of subhaloes in groups at the time when groups enter the cluster. They go on to look at the evolution of these groups by examining which subhaloes remain bound to the group, how this depends on the position and velocity of group members, and when subhaloes become unbound from their group host.

\begin{figure*}
	\includegraphics[width=\textwidth]{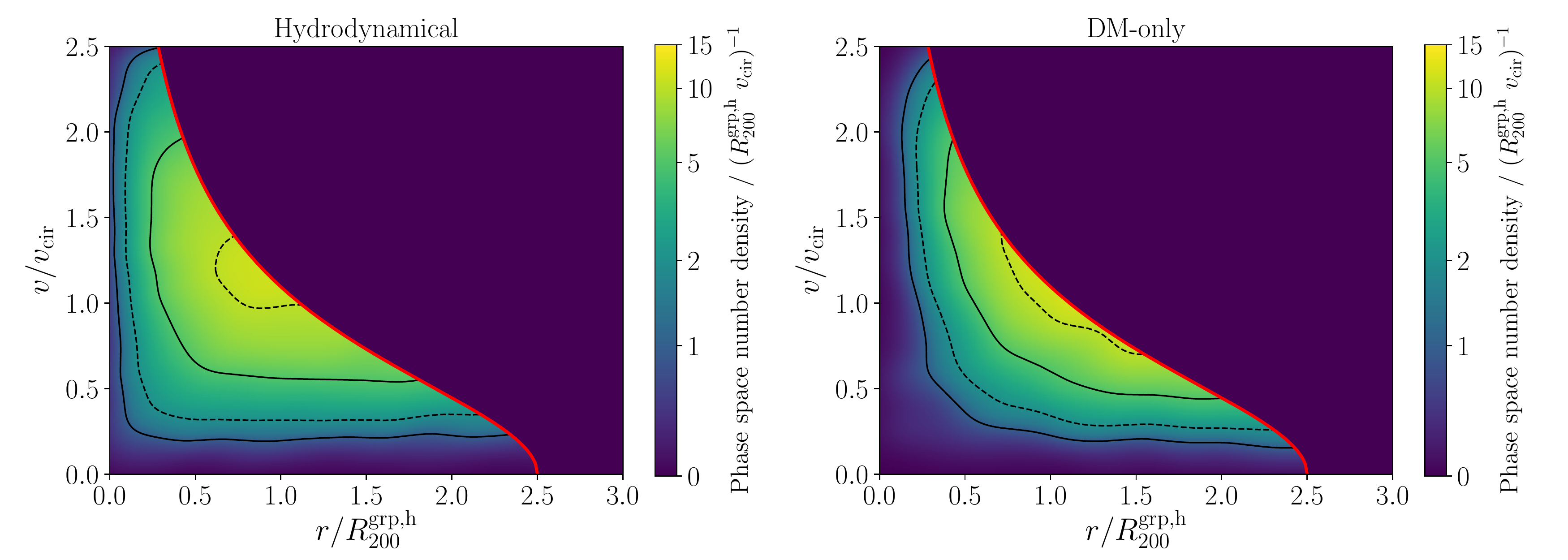}
    \caption{Average phase space distribution of galaxies/subhaloes in groups, at the time of infall, for hydrodynamical (left) and dark matter-only (right) cluster samples. All groups with between five and 50 members are used, across all clusters. Phase space density is the number of galaxies expected in one group, in a square interval of the phase space, of size \mbox{$[R_{200}^{\rm{grp,h}}, v_{\rm{cir}}]$}, where $v_{\rm{cir}}$ is the circular orbital velocity at a distance of $R_{200}^{\rm{grp,h}}$ from the group centre. Contours are at densities of \mbox{$[1,\ 2,\ 5,\ 10]\ ({R_{200}^{\rm{grp,h}}}{v_{\rm{cir}}})^{-1}$} in both plots. Red line shows the boundness criterion given by \Eq{eq:bounded}, meaning that galaxies below this line are bounded to their group host.}
    \label{fig:phasespace}
\end{figure*}

However, as we have shown, the structure of groups in dark matter-only simulations is different to the substructure predicted by more physically motivated hydrodynamical simulations. \Fig{fig:phasespace} demonstrates this further, by showing the phase space distribution of bound satellite galaxies in infalling groups in our sample of 257 dark matter-only (right panel) and hydrodynamical (left panel) clusters described in \Sec{sec:subsample}. We again note that here we are looking at groups at the moment when they enter a galaxy cluster (i.e. the first snapshot at which the group host is inside $R_{200}^{\rm{clus,h}}$), as opposed to \Sec{sec:radial_dens}, in which we look at all groups in the cluster outskirts. The phase space consists of the radial distance of a galaxy from its host group halo, in units of $R_{200}^{\rm{grp,h}}$, and its velocity relative to the group halo, in units of $v_{\rm{cir}}$, the circular orbital velocity at $r=R_{200}^{\rm{grp,h}}$. This figure shows a 2D KDE of the stacked phase space data for all infalling groups across the cluster samples, and is normalised by the total number of infalling groups in each sample. As the typical size of a group ($\sim8$ members) is small, using a KDE with an optimised bandwidth allows the mean distribution of galaxies to be clearly seen.

The phase space of group members in our dark matter-only simulations is in agreement with the distribution found by \citet{choquechallapa2019}, despite their use of lower mass clusters ($\sim10^{14}\ h^{-1}M_{\odot}$), and a lower limit on satellite masses ($\sim10^{7.8}\ h^{-1}M_{\odot}$). The greatest concentration of subhaloes in this phase space is close to the line representing the boundness criterion, spread between approximately \mbox{$0.7R_{200}^{\rm{grp,h}}$} and \mbox{$1.5R_{200}^{\rm{grp,h}}$}, with the maximum located at \mbox{$r=1.1R_{200}^{\rm{grp,h}}$}, \mbox{$v=v_{\rm{cir}}$}. We also find that there are very few subhaloes near to the central regions of groups, particularly with low velocities. Only $9\%$ of dark matter-only groups contain at least one subhalo with \mbox{$r<0.3R_{200}^{\rm{grp,h}}$} and \mbox{$v<v_{\rm{cir}}$} (excluding the host halo at the centre of each group). This is a region that previous work, such as that of \citet{choquechallapa2019}, has also shown to contain few satellites in dark matter-only simulations.

However, carrying out this analysis with the hydrodynamical simulations, as shown by the left panel of \Fig{fig:phasespace}, gives a different distribution of galaxies. The most prominent region of high phase space density does not extend as far from the group centres, as it reaches from \mbox{$0.6R_{200}^{\rm{grp,h}}$} to \mbox{$1.1R_{200}^{\rm{grp,h}}$}, with a maximum at \mbox{$r=0.9R_{200}^{\rm{grp,h}}$}, \mbox{$v=1.2v_{\rm{cir}}$}. The central regions are also more populated with galaxies; $33\%$ of hydrodynamical groups have at least one galaxy in the central, low-velocity region described in the previous paragraph.

\begin{figure}
	\includegraphics[width=\columnwidth]{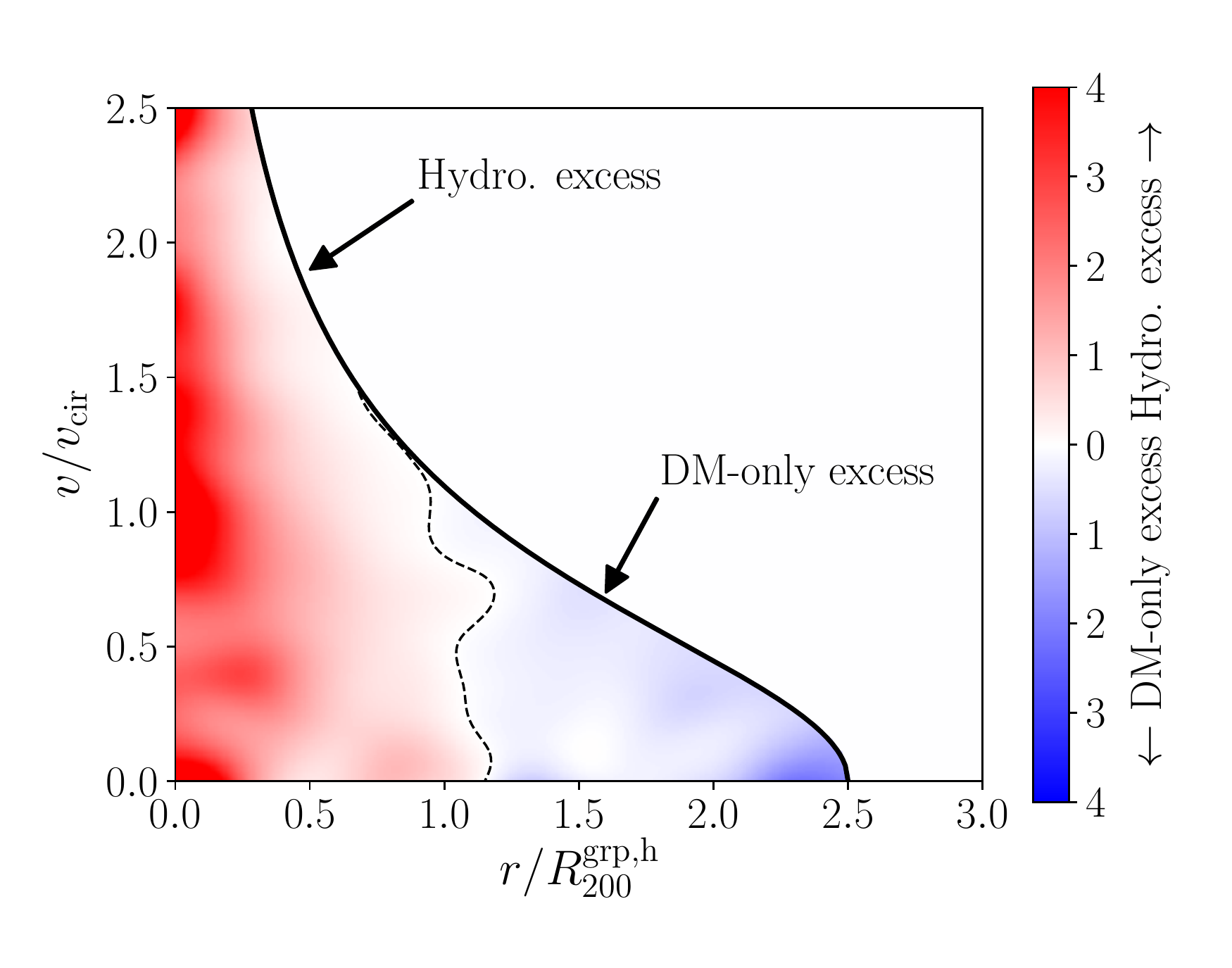}
    \caption{Fractional difference of the phase space distributions of group members in hydrodynamical and dark matter-only simulations, shown in \Fig{fig:phasespace}. The values represent the fractional difference of the greater density value relative to the lower value, and the colour represents whether the hydrodynamical or dark matter-only simulations have an excess of galaxies in this region. These two regimes are separated by the dashed contour at zero, such that hydrodynamical groups have a greater density of galaxies to the left of the line, and dark matter-only groups to the right.}
    \label{fig:phasespace_ratio}
\end{figure}

The differences between the phase spaces are demonstrated more clearly in \Fig{fig:phasespace_ratio}. This plot shows the fractional difference between the phase space density of the hydrodynamical and dark matter-only simulations, represented by the colours in \Fig{fig:phasespace}. In the regions marked as `Hydro. excess' in \Fig{fig:phasespace_ratio}, the colour represents the fractional increase of the hydrodynamical density, relative to the dark matter-only. Similarly, in the `DM-only excess' regions, the quantity plotted is the density excess in the dark matter-only phase space, relative to the hydrodynamical. For example, a value of 1.5 in the `DM-only excess' region would mean that the phase space density in this region is $150\%$ greater in the dark matter-only groups than in the hydrodynamical groups (i.e. $2.5$ times the magnitude). The dashed line marks a contour where the phase space densities are the same, and clearly divides the phase space into two distinct regions. At greater distances from the centres of groups (to the right of this contour), dark matter-only simulations over-predict the abundance of galaxies in this region of phase space. Meanwhile, closer to the centres of groups, dark matter-only simulations under-predict the numbers of galaxies.

\Fig{fig:phasespace_ratio} shows that there are more galaxies beyond \mbox{$1.1R_{200}^{\rm{grp,h}}$} in infalling dark matter-only groups, regardless of the relative velocity of galaxies. This appears to contradict the results of \Sec{sec:radial_dens} and \Fig{fig:dm_hydro_all_groups}, which show that the number density of galaxies in the outskirts of groups is the same in dark matter-only and hydrodynamical simulations. However, because of the larger $R_{200}^{\rm{grp,h}}$ values for dark matter-only groups, these groups have approximately twice the volume of hydrodynamical groups, which will lead to twice as many galaxies in a given radial region, in units of $R_{200}^{\rm{grp,h}}$. Indeed, in most of the phase space with \mbox{$r>R_{200}^{\rm{grp,h}}$}, the relative excess of dark matter-only subhaloes is approximately equal to one.

\Fig{fig:phasespace_ratio} also shows the excess of galaxies in group centres in the hydrodynamical simulations. As discussed earlier in this section, dark matter-only groups are less likely to contain galaxies at very small radii and with low velocities. This is shown by the large hydrodynamical density excess: in the region with \mbox{$r<0.3R_{200}^{\rm{grp,h}}$} and \mbox{$v<v_{\rm{cir}}$}, the average excess is 2.8, corresponding to almost four times as many galaxies in this region of phase space in the hydrodynamical simulations. However, there is a similar excess for all galaxies in these central regions, regardless of their relative velocities.

\subsubsection{Resolution effects}
\label{sec:resolution}

\begin{figure*}
	\includegraphics[width=\textwidth]{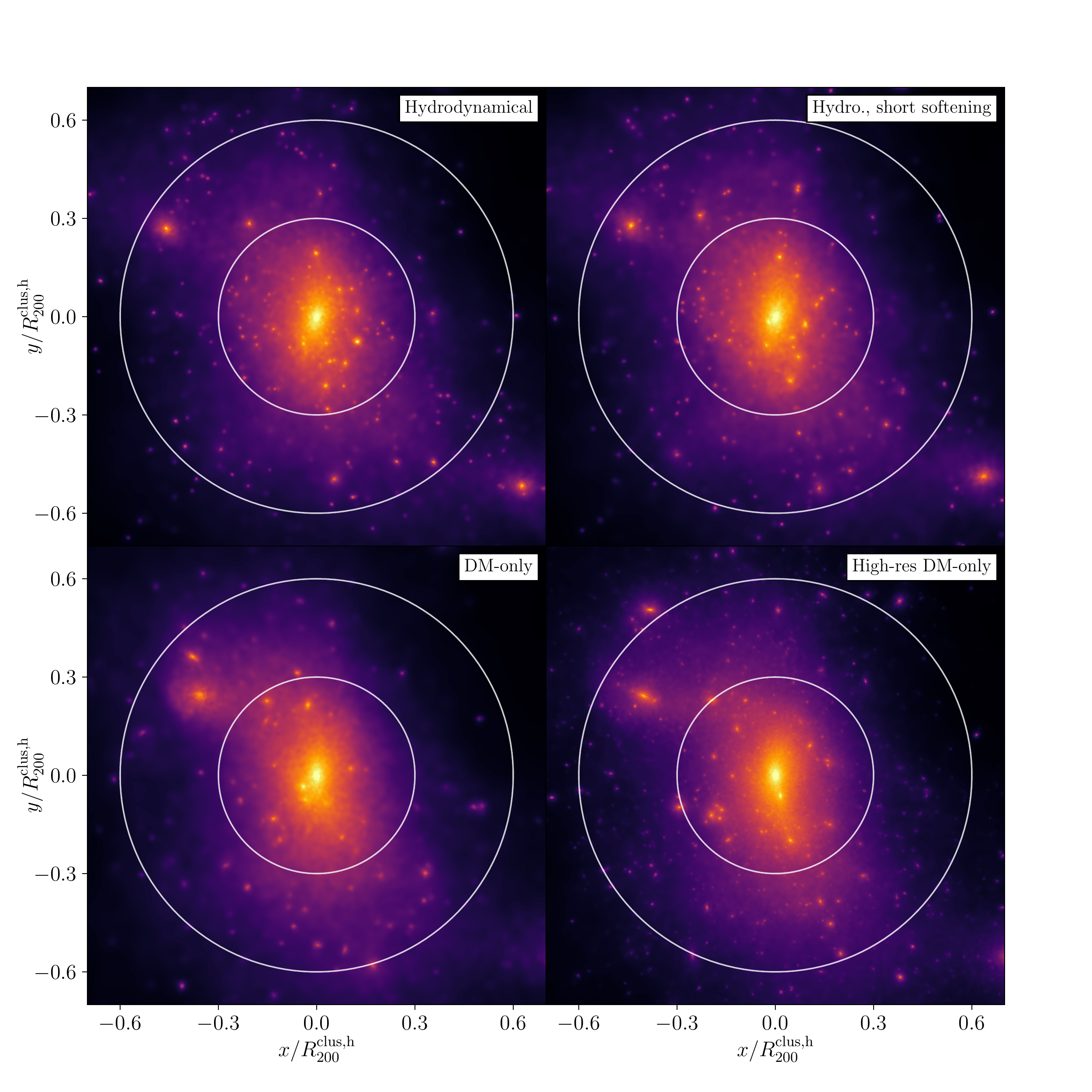}
    \caption{Dark matter distribution of inner region of \mbox{{\sc{cluster}}\_0002}, which was simulated four times. Top-left: the original hydrodynamical run. Top-right: hydrodynamical run with a reduced gravitational softening length for stars. Bottom-left: original dark matter-only run. Bottom-right: high-resolution dark matter-only run. Overlaid circles are at \mbox{$0.3R_{200}^{\rm{clus,h}}$} and \mbox{$0.6R_{200}^{\rm{clus,h}}$} from the cluster centre; the value of $R_{200}^{\rm{clus,h}}$ varies by approximately $1\%$ between the simulations. Note the visibly greater number of compact dark matter haloes within \mbox{$0.3R_{200}^{\rm{clus,h}}$} in the panels representing the hydrodynamical runs.}
    \label{fig:cluster0002}
\end{figure*}

As described in \Sec{sec:hydro} and \Tab{tab:partdata}, we have also simulated one of the clusters (referred to as \mbox{`{\sc{cluster}}\_0002'}) from \threehun\ sample two more times, with a shorter gravitational softening for stellar particles, and in dark matter-only at a resolution eight times greater than the standard dark matter-only clusters. These four simulations of a single cluster, using the same initial conditions and cosmological code, are shown in \Fig{fig:cluster0002}. The top two panels show the dark matter distribution of \mbox{{\sc{cluster}}\_0002} in the two hydrodynamical runs, and the bottom two panels show the dark matter-only simulations.

We find that decreasing the softening length in the hydrodynamical simulation has no significant effect on either the radial profiles in \Sec{sec:radial_dens}, or the phase space of members in infalling groups (\Sec{sec:phase_space}), for this cluster. Conversely, comparing the original and high-resolution dark matter-only runs shows that a change in resolution does have some effect on our results. For example, the radii, $R_{200}^{\rm{grp,h}},$ of the groups in the cluster outskirts are on average $11\%$ smaller in the high-resolution dark matter-only simulation, compared to the original dark matter-only simulation. However, due to the relatively small sample of groups in the outskirts of this one cluster, the difference is not significant ($\sim1\sigma$). 

\Tab{tab:res_substructures} shows the number of galaxies within $0.5R_{200}^{\rm{clus,h}}$ of the centre of the cluster shown in \Fig{fig:cluster0002}, and the total number of galaxies found within $0.5R_{200}^{\rm{grp,h}}$ of group centres in the cluster outskirts, for each of the four simulations. This too is somewhat affected by increasing the resolution; the high-resolution dark matter-only simulation contains more subhaloes in the inner region of the cluster, and more subhaloes in the centres of surrounding groups. However, these increases that are caused by the greater dark matter resolution are less than the difference between the hydrodynamical and dark matter-only simulations. Moreover, the difference in resolution between the hydrodynamical and dark matter-only simulations is much less than a factor of eight (the difference between the standard and high-resolution dark matter-only simulations), so this slight difference in resolution cannot fully account for the differences between the hydrodynamical and dark matter-only simulations seen throughout this work. Finally, we note that slightly more groups are found in the cluster outskirts in the high-resolution dark matter-only simulation. When this is accounted for, the average number of galaxies within $0.5R_{200}^{\rm{grp,h}}$ of the centre of a group is actually unaffected by resolution.

\begin{table}
	\centering
	\caption{Number of galaxies within $0.5R_{200}^{\rm{clus,h}}$ of the centres of the clusters in \Fig{fig:cluster0002}, and total number of galaxies within $0.5R_{200}^{\rm{grp,h}}$ of groups around these clusters. Note that using a high-resolution dark matter-only simulation does increase these quantities relative to the regular dark matter-only simulation, but less so than using hydrodynamical simulations.}
	\label{tab:res_substructures}
	\begin{tabular}{ccc} 
		\hline
		Simulation & $N(r<0.5R_{200}^{\rm{clus,h}})$ & $N(r<0.5R_{200}^{\rm{grp,h}})$ \\
		\hline
		Hydrodynamical & 264 & 35\\
		Reduced softening & 258 & 35\\
		Dark matter-only & 86 & 14\\
		High-res DM-only & 198 & 23\\
		\hline
	\end{tabular}
\end{table}

This demonstrates that our findings are an effect of the inclusion of baryonic material, rather than a resolution effect, and the differences between the four simulations of this same cluster are in fact visible in \Fig{fig:cluster0002}. The main features of the cluster are similar in each of the simulations, although they show some variation, as these plots focus on the central cores of the clusters \mbox{($r<0.7R_{200}^{\rm{clus,h}}$)}. Nevertheless, there are visibly more compact dark matter haloes in the central regions of the top two panels (showing the hydrodynamical runs), particularly very close to the cluster centres. There is very little difference in the substructure produced when increasing the resolution or decreasing the softening length.

\section{Discussion \& implications}
\label{sec:discussion}

The difference in the distribution of galaxies between hydrodynamical and dark matter-only simulations clearly has implications for the study of galaxy groups and clusters via simulations. This difference could have three main causes. The inclusion of gas and baryonic physics in the simulations may result in different substructure forming, both within the cluster itself, and in groups in the cluster outskirts. Alternatively the difference could be due to systematic issues with the halo finder (in our case, {\sc{ahf}}), which then impact the halo catalogue that it produces. Finally, the presence of baryonic material may alter the properties of dark matter haloes, making them more likely to survive in certain environments.

Tidal stripping can cause the mass of a dark matter halo to decrease on entering a larger halo, whilst having a more minor effect on the baryonic material at the centre of such haloes \citep{smith2016}. A future study (Mostoghiu et al. (in prep.)) will investigate this using \threehun\ simulations. Previous work has also indicated that the ratio of stellar mass to halo mass is greater for backsplash galaxies, which have experienced the environmental effects of a cluster in their past, indicating the stripping of their dark matter haloes \citep{knebe2011a, haggar2020}. Furthermore, tidal effects were partly responsible for the over-merging problem, as seen in early simulations \citep{moore1998}. If these tidal effects are enhanced in dark matter-only simulations relative to hydrodynamical simulations, this could result in a drop in the number density of galaxies in denser regions, causing the deficit of galaxies seen in the centres of groups and clusters. 

However, there are a number of issues with this. As tidal effects are a result of the gravitational potential of host groups and clusters, they are present in hydrodynamical simulations, as well as other stripping mechanisms such as ram pressure stripping \citep{arthur2019}. This would cause gas to be stripped from galaxies in groups and clusters, as well as dark matter, potentially leaving a fully stripped galaxy core that consists mostly of stellar material. Such objects would not be found in a dark matter-only simulation, and so would indeed lead to a deficit of galaxies. However, as described in \Sec{sec:galgroups}, we remove all galaxies from the hydrodynamical simulations that have more than $30\%$ of their mass contained in stellar particles, corresponding to about $1\%$ of all objects within $5R_{200}^{\rm{clus,h}}$ of the clusters. This would remove these heavily stripped galaxies, meaning that they would be absent in both the hydrodynamical and dark matter-only simulations. 

Furthermore, we note that the deficit in subhaloes in dark matter-only groups is similar at all velocities in the group centre. This includes high-velocity galaxies which are likely near pericentre of a radial orbit, moving quickly from the group outskirts into its dense central region, as well as low-velocity galaxies, which are on roughly circular, bound orbits near the middle of the group. Assuming an inside-out formation history of groups \citep{vanderburg2015} implies that these low-velocity galaxies joined their host group at an earlier time in its history, when it was less massive, and have since settled into virialised orbits. Consequently, we would expect these low-velocity galaxies to be less affected by tidal effects than those entering the group at a later time. This is not the case, as we see that galaxies are more prevalent in the hydrodynamical simulations, regardless of their velocities. Additionally, the right panel of \Fig{fig:phasespace} shows that very low-velocity central galaxies (\mbox{$r<0.3R_{200}^{\rm{grp,h}}$}, \mbox{$v<0.5v_{\rm{cir}}$}) are almost completely absent in dark matter-only simulations, despite these galaxies being less affected by tidal effects. This indicates that tidal effects are not playing a strong role in changing the number density of galaxies in group and cluster centres. However, the large deficit of extremely high-velocity haloes ($v\gtrsim2v_{\rm{cir}}$) in the dark matter-only simulations could indeed be a result of tidal effects. Similarly, the (small) deficit of backsplash galaxies around our dark matter-only clusters could also be due to tidal effects, as backsplash galaxies are known to follow highly radial orbits and so experience strong tidal forces \citep{knebe2020}.

In spite of this, previous work has found results analogous to ours, that have been largely attributed to tidal forces. \citet{libeskind2010} compare two simulations of a system similar to the Local Group, simulated from the same initial conditions in both dark matter-only, and with full hydrodynamics. They too find that subhaloes are more concentrated in the centres of host haloes in their hydrodynamical simulation. The reason provided for this is that the dense baryonic region at the centre of hydrodynamical haloes restricts the tidal stripping of the dark matter halo. This leads to stronger dynamical friction in the hydrodynamical simulation, resulting in these objects being dragged towards the centre of their host, and increasing the galaxy number density. While this may partially explain the results in our work, it would also lead to a drop in the total masses of galaxy haloes in the dark matter-only simulations. In fact, we find that the cumulative halo mass functions for our hydrodynamical and dark matter-only clusters do not differ by more than $25\%$, between \mbox{$M_{200}=10^{10.5}\ h^{-1}M_{\odot}$} and \mbox{$M_{200}=10^{15}\ h^{-1}M_{\odot}$}. This indicates that, although the mechanism described by \citet{libeskind2010} may partially contribute to the results in this work, the effect does not seem to be strong enough to fully explain the differences in density that we find in groups and clusters.

An alternative explanation for the trends seen in this work is a numerical effect, that the inclusion of baryonic material will alter the effectiveness of halo finders that are used in simulations. Most galaxy halo finders are based off of one of two principles; typically, they detect the centres of galaxy haloes by either searching for local peaks in a density field, or by finding groups of particles that are close together, either in physical space or phase space (see \citet{knebe2011b} for a far more extensive overview). Both of these general methods rely on locating a dense halo centre, and then expanding from this to determine the extent of the galaxy halo.

Consequently, the presence of a dense region of star and gas particles at the middle of a dark matter halo will benefit both of these types of halo finder, as this will help a halo satisfy the conditions to become a seed for the halo finder to use. However, this non-physical explanation is unlikely to be the only cause of the difference between the simulations. Our third explanation, that the dense central region of baryonic material can also have a physical effect on the dark matter, is strongly supported by both this work and previous studies in the literature. As described by \citet{blumenthal1986} and \citet{vandaalen2011}, this dense baryonic region can increase the steepness of the dark matter density profile in a halo centre. This would indeed enhance the ability of a halo finder to detect the halo, but because of a physical difference in the simulations, not simply a numerical effect. The initial step in {\sc{ahf}} involves using a refined grid to locate peaks in the density field \citep{knollmann2009}, and so if these peaks are less sharp in a dark matter-only simulation, detection of galaxies near to the centres of groups will be more challenging. These subhaloes may instead be included as part of the group halo, potentially contributing to the greater sizes of group haloes in dark matter-only simulations, as shown in \Fig{fig:r200_hist}. 

We stress that, due to the nature of most halo finders, this effect is not unique to the halo finder used in this work. The visible differences between the simulations in \Fig{fig:cluster0002} show that the baryonic material is also having a physical impact on the dark matter, and so this is not purely a numerical effect. Similar effects would likely be observed with many other halo finders used widely in cosmological simulations. For example, \citet{knebe2011b} examine 16 different halo finders in a dark matter-only cosmological volume, and find that the cumulative halo mass functions predicted by these agree to within $\sim10\%$ over nearly four orders of magnitude, from \mbox{$M_{200}=2\times10^{11}\ h^{-1}M_{\odot}$} to \mbox{$M_{200}=10^{15}\ h^{-1}M_{\odot}$}. \citet{onions2012} find a similar result in a lower mass regime, instead studying the effectiveness of 11 halo finders at detecting subhaloes within a Milky Way-sized dark matter halo. They too find a variation of approximately $10\%$ in the cumulative halo mass function, this time between \mbox{$M_{200}=6\times10^{6}\ h^{-1}M_{\odot}$} and \mbox{$M_{200}=10^{10}\ h^{-1}M_{\odot}$}. They also demonstrate that this result still holds in the dense, central regions of the galaxy halo.

For the specific example of infalling galaxy groups that we have investigated in \Sec{sec:phase_space}, we show that the tools typically used to analyse simulation data lead to different views of the composition of galaxy groups depending on the inclusion of baryonic material, which is not immediately obvious. This difference in the composition of groups will affect conclusions relating to their evolution. For example, \citet{choquechallapa2019} determine, amongst other results, the fraction of galaxies that become unbound from infalling groups in dark matter-only simulations, and how this depends on the position/velocity of galaxies relative to their group host. However, our work adds a caveat to results such as these. We show that dark matter-only simulations underestimate the fraction of central, low-velocity galaxies, which are more likely to remain bound to infalling groups. We plan to build on the work in \Sec{sec:phase_space} in a follow-up paper, using hydrodynamical simulations to investigate the infall of these galaxy groups, study how the dynamical properties of their constituent galaxies change during infall, and how this depends on the properties of the group.

\section{Conclusions}
\label{sec:conclusions}

In this work we have examined the differences in galaxy cluster substructure produced by hydrodynamical and dark matter-only simulations. We make this comparison by using a suite of hydrodynamical and dark matter-only simulations, to obtain two cluster samples that use the same cosmology, initial conditions, simulation codes and analysis. We then use the specific example of the phase space of infalling galaxy groups to investigate how the analysis of cluster simulations could be affected by these differences. Our findings are summarised below.

\begin{itemize}

\item Apart from the outskirts of galaxy groups and clusters, where the number density of galaxies is below \mbox{$\sim1$ Mpc$^{-3}$}, dark matter-only simulations underestimate the radial number density profiles of galaxies in clusters, and in groups located in cluster outskirts. It is only at distances beyond \mbox{$\sim2R_{200}$} from the centres of groups/clusters that the profiles are indistinguishable. 

\item Closer to the centres of groups and clusters, the deficit of galaxies in dark matter-only simulations increases. At \mbox{$r=0.1R_{200}$} in clusters, the number density of galaxies is four times greater in hydrodynamical simulations. At the same distance in groups (scaled by $R_{200}$), the average number density is 10 times greater in hydrodynamical simulations.

\item In galaxy groups that are entering a cluster, the deficit of galaxies in dark matter-only simulations is particularly pronounced when considering galaxies close to the group centre, with either high or low velocities relative to their group host. In some regions of the position-velocity phase space, there are up to five times as many galaxies in an average hydrodynamical group. 

\item The presence of a dense region of baryonic material in the centres of hydrodynamical haloes, and the increased central density of dark matter caused by this, means that galaxies produce a sharper peak in the density field within hydrodynamical simulations. The increased prominence of over-densities, and hence the increased contrast of galaxies against their host group, makes them easier for halo finders to detect in hydrodynamical simulations. The consequences of this are the discrepancies between the two simulation types that we describe in this work.

\end{itemize}

The results from \Sec{sec:phase_space} show that infalling galaxy groups appear less compact when using dark matter-only simulations, compared to more physically motivated hydrodynamical simulations. This will affect the evolution of these groups, and the fate of group members after the infall of their host group, as previous work has shown that compact groups appear more likely to survive cluster infall \citep{choquechallapa2019}. We will investigate this, using hydrodynamical simulations, in a follow-up paper.

However, this work has wider implications for cosmological simulations. We show that the use of dark matter-only simulations, either as an approximation of a full-physics simulation, or as a framework for techniques such as semi-analytic modelling or halo occupation models, may need to be adjusted. Many semi-analytic models already account for similar effects, by including `orphan galaxies' in their catalogues; these are subhaloes that have been heavily stripped or disrupted as they approach the centre of their host halo, and so can no longer be located in the simulation \citep{contini2015,pujol2017,cora2018}. Similar methods have been used in halo occupation models, to find orphan galaxies and to adequately populate clusters with satellite galaxies \citep{carretero2015,guo2016}.

Despite this, such methods are not widely used in dark matter-only simulations, and as we show, this can potentially lead to different conclusions regarding the structure and composition of groups and clusters. As a minimum, the caveat that dark matter-only simulations produce incomplete halo catalogues in dense cosmological regions must be included. Further to this, corrections need to be made to work involving dark matter-only simulations, to account for the fact that observational surveys of groups and clusters will find a greater population of central galaxies than predicted by these simulations.

\section*{Acknowledgements}

We thank the anonymous referee for their helpful comments and suggestions, which have helped to make this work clearer and more rigorous.

This work has been made possible by \threehun\ collaboration\footnote{\url{https://www.the300-project.org}}. This work has received financial support from the European Union's Horizon 2020 Research and Innovation programme under the Marie Sk\l{}odowskaw-Curie grant agreement number 734374, i.e. the LACEGAL project\footnote{\url{https://cordis.europa.eu/project/rcn/207630\_en.html}}. \threehun\ simulations used in this paper have been performed in the MareNostrum Supercomputer at the Barcelona Supercomputing Center, thanks to CPU time granted by the Red Espa\~nola de Supercomputaci\'on.

RH acknowledges support from STFC through a studentship. He also thanks Liza Sazonova for valuable discussions relating to the analysis in this work, and Andrew Benson for his helpful interpretation of these results. AK and GY are supported by the \textit{Ministerio de Econom\'ia y Competitividad} and the \textit{Fondo Europeo de Desarrollo Regional} (MINECO/FEDER, Spain) under research grant PGC2018-094975-C21. AK further acknowledges support from the Spanish Red Consolider MultiDark FPA2017-90566-REDC and thanks Beastie Boys for check your head.

This work makes use of {\sc{Astropy}} \citep{robitaille2013}, a community-developed core {\sc{Python}}\footnote{\url{https://www.python.org}} package for astronomy, as well as the {\sc{SciPy}} \citep{virtanen2020}, {\sc{NumPy}} \citep{vanderwalt2011}, {\sc{Matplotlib}} \citep{hunter2007}, {\sc{pandas}} \citep{mckinney2010}, {\sc{Py-SPHViewer}} \citep{benitez-llambay2015} and {\sc{PyQt-Fit}}\footnote{\url{https://pypi.org/project/PyQt-Fit/}} packages.

The authors contributed to this paper in the following ways: RH, MEG, FRP and AK formed the core team. RH analysed the data, produced the plots and wrote the paper along with MEG and FRP. AK produced the halo catalogues and merger trees. GY supplied \threehun\ simulation data, and the dark matter-only data. All authors had the opportunity to comment on the paper.

\section*{Data availability}

The data underlying this work has been provided by \threehun\ collaboration. The data may be shared on reasonable request to the corresponding author, with the permission of the collaboration.



\bibliographystyle{mnras}
\bibliography{hydro_dm_grps} 



\appendix

\section{Dependence of number density profile on group mass}
\label{sec:appendix_grp_mass}

In \Sec{sec:masses_radii}, we briefly discuss the effect of group host halo mass on the radial number density of galaxies within the group, for groups in the outskirts of a cluster (as shown in \Fig{fig:dm_hydro_all_groups}).

In both the hydrodynamical and dark matter-only simulations, the range of group halo masses is approximately two orders of magnitude. To investigate the effect of group mass on the radial density profiles of the groups, we split each sample of groups into three mass bins, each containing approximately equal numbers of groups. We then compare the radial number density profiles of groups in each bin, for the hydrodynamical and dark matter-only simulations. 

\begin{figure*}
	\includegraphics[width=\textwidth]{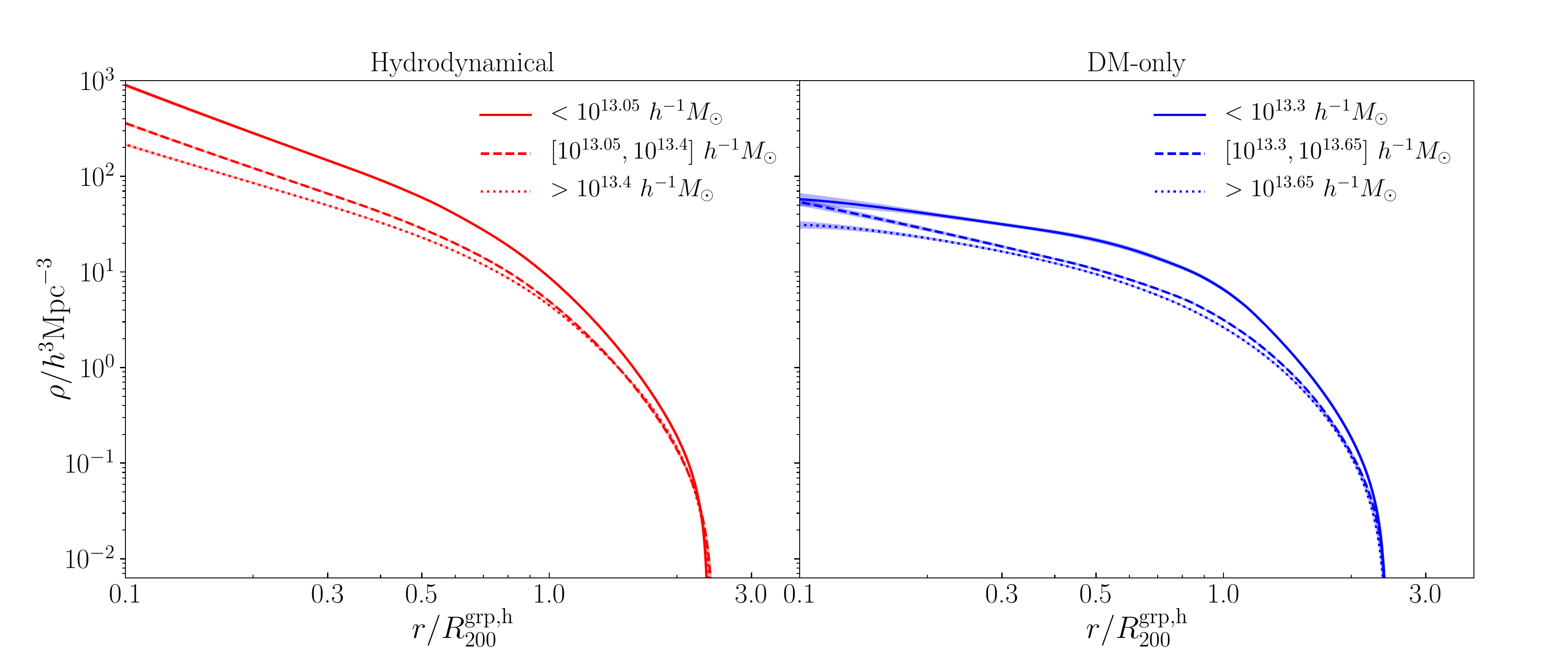}
    \caption{Radial number density of galaxies in groups (as in \Fig{fig:dm_hydro_all_groups}), for hydrodynamical and dark matter-only simulations, split by group host halo mass. For each class of cluster simulations, we split the groups into three mass bins, with approximately equal numbers of groups in each bin; these mass bins are shown in the legend. Shaded regions represent the uncertainty in the average density profile. For clarity, the spread of the data for each sample of groups is not shown.}
    \label{fig:dm_hydro_groups_mbins}
\end{figure*}

This is shown in \Fig{fig:dm_hydro_groups_mbins}. We find that the difference in radial number density between the high-mass and low-mass bins is less than a factor of 2.5 at all radii in the dark matter-only simulations. In the hydrodynamical simulations, the maximum difference between the two mass bins is a factor of four, but only in the very centres of the groups. This is a significant difference, but is less than the difference between the hydrodynamical and dark matter-only simulations, which reaches a maximum of a factor of 10. It is therefore unlikely to affect the conclusions of this work.

\begin{figure*}
	\includegraphics[width=\textwidth]{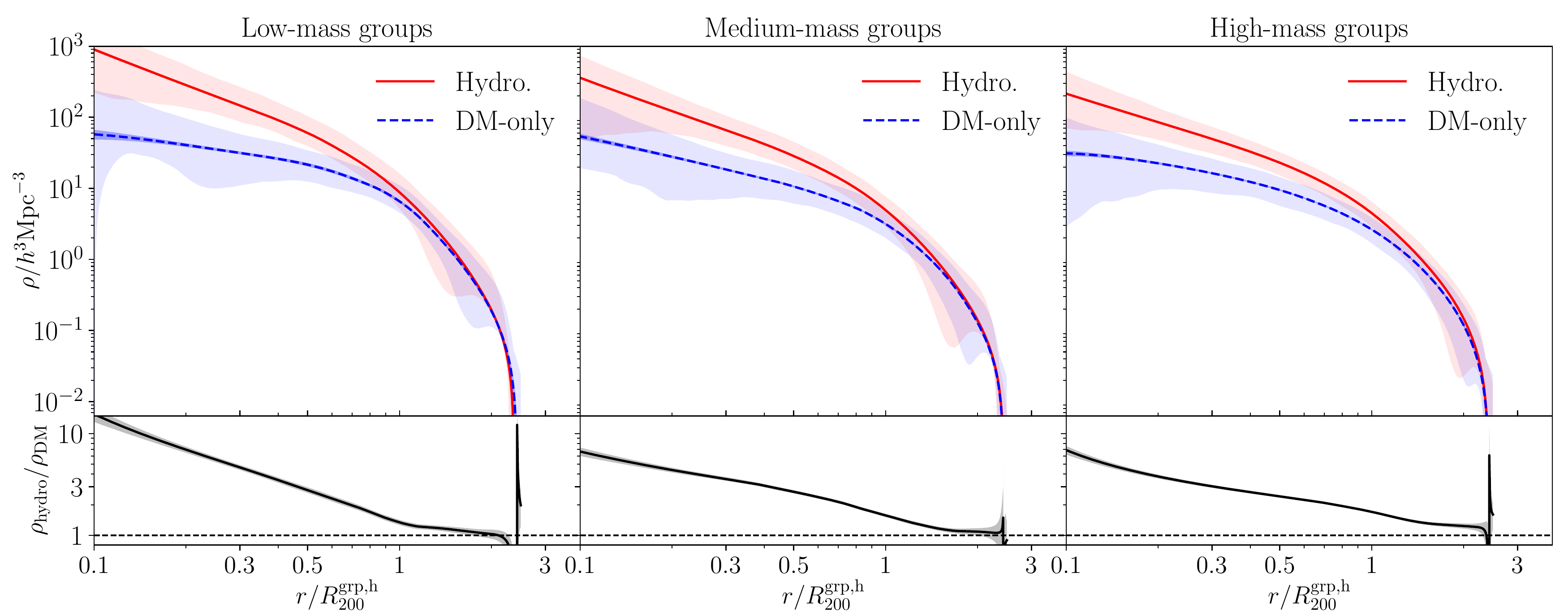}
    \caption{Radial number density of galaxies in groups located in cluster outskirts at $z=0$ (top panels). These plots show the same data as \Fig{fig:dm_hydro_all_groups}, but split into three mass bins, each containing approximately one third of the groups in the radial range \mbox{$[R_{200}^{\rm{clus,h}}, 5R_{200}^{\rm{clus,h}}]$} around a cluster. For hydrodynamical simulations, the low, medium and high-mass groups are those with a group host halo mass, $M_{200}^{\rm{grp,h}}$, in the ranges \mbox{($<10^{13.05}\ h^{-1}M_{\odot}$)}, \mbox{($[10^{13.05},10^{13.4}]\ h^{-1}M_{\odot}$)} and \mbox{($>10^{13.4}\ h^{-1}M_{\odot}$)}, respectively, as shown in \Fig{fig:dm_hydro_groups_mbins}. For dark matter-only simulations, the mass bins are \mbox{($<10^{13.3}\ h^{-1}M_{\odot}$)}, \mbox{($[10^{13.3},10^{13.65}]\ h^{-1}M_{\odot}$)} and \mbox{($>10^{13.65}\ h^{-1}M_{\odot}$)}. Shaded regions represent $1\sigma$ uncertainty in the mean radial number density profile, although these are mostly too small to be seen. Bottom panel shows fractional residuals (the ratio of the hydrodynamical and dark matter-only profiles, as defined in \Sec{sec:radial_dens}).}
    \label{fig:dm_hydro_groups_mbins_ratio}
\end{figure*}

Most importantly, when comparing mass bins between the simulations, the same trend is seen as in \Fig{fig:dm_hydro_all_groups}, for each of the three mass bins. As \Fig{fig:dm_hydro_groups_mbins_ratio} shows, when considering either low, medium or high-mass groups in the cluster outskirts, the number density of galaxies, $\rho$, is consistently greater in the hydrodynamical simulations. The difference is slightly greater in the smallest groups, but even in groups with large masses, the number density of galaxies in the centre of hydrodynamical groups is seven times greater than in dark matter-only groups. This is shown by the fractional residual at the bottom of each plot -- these are calculated in the same was is described in \Sec{sec:radial_dens}. The same trend is seen when comparing like-for-like mass bins (i.e. when comparing groups with halo masses in the range \mbox{$[10^{13.0}, 10^{13.5}]\ h^{-1}M_{\odot}$} in both the hydrodynamical and dark matter-only simulations).

This shows that the mass of a group does impact the number density of galaxies within the group. However, it also demonstrates that this effect is substantially smaller than the difference caused by the inclusion of baryonic material within the simulations.


\bsp	
\label{lastpage}
\end{document}